\documentclass[showpacs, aps, physics,reprint,amsmath, amssymb, mathtools,pra,floatfix
]{revtex4-2}
\usepackage{upgreek}
\usepackage{xcolor}
\usepackage{bbold}
\usepackage{multirow}
\usepackage{dsfont}
\usepackage{suffix}
\usepackage{mathtools}
\usepackage{graphicx}% Include figure files
\usepackage{dcolumn}% Align table columns on decimal point
\usepackage{bm}
\usepackage{float}

\begin{document}
%\linenumbers
\preprint{APS/123-QED}

\title{Spectral crossovers in non-Hermitian spin chains :\\ comparison with random matrix theory }% Force line breaks with \\
%\thanks{A footnote to the article title}%

\author{Ayana Sarkar}
\email{ayanas1994@gmail.com, (she/her)}
\author{Sunidhi Sen}
\email{sensunidhi96@gmail.com, (she/her)}
\author{Santosh Kumar}%
 \email{skumar.physics@gmail.com, (he/him)}
\affiliation{Department of Physics, Shiv Nadar Institution of Eminence, Gautam Buddha Nagar, Uttar Pradesh-201314, India \\}%

\date{\today}% It is always \today, today,
             %  but any date may be explicitly specified

\begin{abstract}
We present a systematic investigation of the short-range spectral fluctuation properties of three non-Hermitian spin chain Hamiltonians using complex spacing ratios (CSR). Specifically, we focus on the non-Hermitian variants of the standard one-dimensional anisotropic XY model having intrinsic rotation-time ({$\mathcal{RT}$}) symmetry that has been explored analytically by Zhang and Song [Phys. Rev. A {\bf 87}, 012114 (2013)]. The corresponding Hermitian counterpart is also exactly solvable and has been widely employed as a toy model in several condensed matter physics problems. We show that the presence of a random field along the $x$-direction together with the one along $z$-direction facilitates integrability and $\mathcal{RT}$-symmetry breaking leads to the emergence of quantum chaotic behaviour. This is evidenced by a spectral crossover closely resembling the transition from Poissonian to Ginibre Unitary Ensemble (GinUE) statistics of random matrix theory. Additionally, we consider two phenomenological random matrix models in this paper to examine 1D-Poisson to GinUE and 2D-Poisson to GinUE crossovers and the associated signatures in CSR. Here 1D and 2D Poisson correspond to real and complex uncorrelated levels, respectively. These crossovers reasonably capture spectral fluctuations observed in the spin-chain systems within a certain range of parameters.
%We also find that for the GinOE to GinUE crossover, while there are some noticeable effects in the CSR due to the presence of real eigenvalues in the GOE limit, with increasing dimension and with a slight perturbation towards GinUE, these are washed out. Therefore, CSR proves inadequate for studying the GinOE-to-GinUE crossover. This observation aligns with the well-established fact that all three classical Ginibre ensembles exhibit identical universal local spectral fluctuations.
\end{abstract}

%\pacs{Valid PACS appear here}% PACS, the Physics and Astronomy
                             % Classification Scheme.
%\keywords{Suggested keywords}%Use showkeys class option if keyword
                              %display desired
\maketitle

%\tableofcontents
\section{Introduction}
\label{Intro}

Non-Hermitian hamiltonians possessing complex eigenvalues arise in a wide variety of systems, including those with dissipation, such as the dissipative kicked rotor, open quantum systems like boundary-driven spin chains with spin injection and ejection terms, and gain-loss Hamiltonians commonly encountered in quantum optics. Among non-Hermitian hamiltonians, a subclass is formed by those which possess the constraint of Parity--Time ($\mathcal{PT}$) or the more general Rotation--Time ($\mathcal{RT}$) symmetry and can exhibit real spectra. These hamiltonians have received a lot of attention since real eigenvalues guarantee unitary time evolution leading to conservation of probability amplitude which is fundamental to describing a quantum theory useful in physical interpretation of natural phenomena. The subject of $\mathcal{PT}$--symmetric quantum mechanics has been deeply enriched by the seminal works of Bender, Mostafazadeh and others~\cite{BB1998,BBM1999,DDT2001,BBJ2002,BBJ2003,Fring2013,Benderbook2019,BBJ2002,Mos2002a,Mos2002b,MB2004,Mos2003,Jon2005}, who have established it as an extension of the conventional or Hermitian quantum mechanics.

Mathematically, the linear parity operator $\mathcal{P}$ performs spatial reflection and has the effect $p \rightarrow -p$ and $x \rightarrow -x$, whereas the anti-linear time-reversal operator $\mathcal{T}$ has the effect of transforming $p \rightarrow -p$, $x \rightarrow x$ and $i \rightarrow -i$. The joint action of $\mathcal{PT}$ together is basically a reflection i.e. $\mathcal{PT} = (\mathcal{PT})^{-1}$. Bender and his co-workers have established an extensive collection of non-Hermitian $\mathcal{PT}$--symmetric Hamiltonians in their research. In general, it has been shown that the reality of quantum spectrum is an outcome of unbroken $\mathcal{PT}$ symmetry. A hamiltonian is called $\mathcal{PT}$ symmetric if its eigenfunctions are simultaneously eigenfunctions of the $\mathcal{PT}$ operator, and in such cases $\mathcal{PT}$ symmetry is not spontaneously broken. Many examples of non-Hermitian hamiltonians possessing $\mathcal{PT}$ symmetry have been discussed in the Refs.~\cite{BB1998,BBM1999,BBJ2002,BBJ2003,Wu1959,Holl1992,Fish1978,Card1985,CM1989,Zam1991,BFM1978,HJT1980,GSZ2005,GSS2007,FF2007,HN1997,NS1998,Bender2007}. Beyond complex hamiltonians with $\mathcal{PT}$ symmetry and possessing real spectra, one comes across hamiltonians which are $\mathcal{RT}$-symmetric. It has been demonstrated in Ref.~\cite{LCKB2020} that $\mathcal{RT}$ symmetry is a \emph{superset}, i.e. a more general notion compared to $\mathcal{PT}$ symmetry, such that a wide class of hamiltonians may be identified that have properties similar to that of $\mathcal{PT}$-symmetric systems, despite not being explicitly $\mathcal{PT}$-symmetric ($\mathcal{PT}$ symmetry is only a special class of $\mathcal{RT}$ symmetry).

In numerous scenarios, non-Hermiticity can be imparted to a system without necessitating its openness in terms of interaction with an external bath or environment. One approach, for instance, involves introducing external imaginary fields to establish $\mathcal{PT}$ symmetry and achieve a real spectrum. Several popular Hermitian spin chain models have been modified to include imaginary interactions, resulting in the emergence of complex spectra in general. For instance, the quantum Ising model in presence of a magnetic field in the $z$-direction as well as an imaginary field in the $x$-direction has been studied analytically in Ref.~\cite{AF2009}. Therein the authors examine various symmetries of the system and study the spin system in the light of perturbation theory, providing some exact results for magnetization along the $z$ and $x$ directions. This modified Ising model is the discretized lattice version of the Yang-Lee model considered by von Gehlen in Refs.~\cite{Geh1991,Geh1994}. In fact the Yang-Lee zeros have recently been observed by measuring the quantum coherence of a probe spin coupled to an Ising-type spin bath. The quantum evolution of the probe spin introduces a complex phase factor which effectively realizes an imaginary magnetic field. It also substantiates that imaginary magnetic fields are not unnatural and are indeed experimentally accomplishable~\cite{PZWCDL2015}. The classical Heisenberg spin chain with $\mathcal{PT}$ symmetry has been studied under the action of Slonczewski spin-transfer torque modeled by applying an imaginary magnetic field~\cite{GV2018}. Interestingly this correlation between the imaginary magnetic field on spin dynamics and Slonczewski spin-transfer torque allows an experimental verification of the $\mathcal{PT}$ symmetry-breaking phase transition in some spin chains~\cite{PLLTRB2012,PLTRB2012}. Exact solutions using the Bethe ansatz technique have been provided for an XXZ spin chain in the presence of an imaginary magnetic field at the boundary in Ref.~\cite{Buca2020}. Such exact solutions are also available for the one-dimensional dissipative Hubbard model with two-body loss \cite{NKU2021}. Giorgi in Ref.~\cite{Gio2010} has studied spontaneous $\mathcal{PT}$-symmetry breaking in an exactly solvable non-Hermitian dimerized chain where non-Hermiticity is introduced via a staggered magnetic field. In Ref.~\cite{ZJS2012} a $\mathcal{PT}$-symmetric non-Hermitian version of a quantum network, originally proposed in Refs.~\cite{CDEL2004,CDDEKL2005}, has been studied in the context of quantum state transfer. Besides the above examples, non-Hermiticity may also arise from $\mathcal{PT}$-symmetric on-site imaginary potentials in tight-binding models and strongly correlated systems~\cite{BFKS2009,JSBS2010,JS2012abc,Z2011abc,Lon2010abc,JS2011,KW2007,ZHLL2011,DSB2011,Yes2011,BZ2011,HJLS2012}. Additionally, some authors have used the $\mathcal{PT}$-symmetric nature of non-Hermitian hamiltonians to produce efficient algorithms to compute their spectra with arbitrarily high precision~\cite{NLJ2013,NLSJ2017}. Studies of $\mathcal{RT}$-symmetric bosonic and fermionic systems have been conducted in diverse capacities, encompassing exactly solvable models, quantum batteries, detection of exceptional points through dynamics, and the identification of unbroken phases using quantum-information-related techniques, in Refs.~\cite{ZS2013,ZS2013b,LCKB2020,LSD2021,KLSD2022,AKLSD2022}.
Some authors have focused on solvable fermionic spin chains. For example, in Refs.~\cite{ZS2013,ZS2013b}, Zhang and Song have analytically studied and identified exceptional points and regions of broken as well as unbroken symmetries in an one-dimensional anisotropic non-Hermitian XY model in transverse magnetic field ($z$-direction) having intrinsic $\mathcal{RT}$-symmetry, with respect to certain parameters.

Despite such extensive analytical explorations, non-Hermitian spin chains of the aforementioned nature have received relatively little attention within the context of random matrix theory (RMT) and quantum chaos. In addition to other objectives, our aim in this paper is to contribute in this direction by studying the short-range fluctuation properties of the anisotropic XY model in a transverse magnetic field ($z$-direction) having intrinsic $\mathcal{RT}$-symmetry, as introduced by Zhang and Song in Ref.~\cite{ZS2013}. In particular, we examine the spectral fluctuations using the statistics of complex spacing ratios (CSR), which constitutes a relatively new metric to deal with complex spectra. Along with the original model, we consider a modification in this hamiltonian by adding a longitudinal magnetic field along the $x$-direction, which breaks its $\mathcal{RT}$-invariant nature. On addition of this field, an integrability to quantum-chaotic transition is observed, portrayed by a symmetry crossover from Poisson to Ginibre Unitary Ensemble (GinUE)-resembling statistics of RMT. We also examine another modification of this spin chain by making the transverse field imaginary, while maintaining the longitudinal magnetic field, and find that the agreement with GinUE improves. These are the some of the key highlights of this work. Furthermore, we consider phenomenological random matrix models to study 1D-Poisson to GinUE and 2D-Poisson to GinUE crossovers. By 1D and 2D Poisson we mean real and complex uncorrelated levels, respectively. This kind of parameter-dependent spectral crossover in Hermitian systems, especially across symmetry classes like Poisson, Gaussian orthogonal ensemble (GOE), Gaussian unitary ensemble (GUE), and Gaussian symplectic ensemble (GSE) have been observed and exhaustively studied in several many body quantum systems such as spin chains, lattice models, periodically driven systems, gas of interacting particles etc ~\cite{RNM2004,MM2014,ARB2002,KD2004,SM2016,MMR2014,RN2016,LLA2015,ZPP2008,Yur2023a,Yur2023b,SMW2017a,SMW2017b,SMW2017c,KKSG2022,KKSG2023}. However, when it comes to non-Hermitian systems, there have been fewer studies related to parameter-dependent Poisson and GinUE-like behavior, as well as spectral transitions~\cite{SRP2020,GGK2022}. With this in mind, in this study, we investigate the integrability-to-quantum-chaotic crossovers in non-Hermitian spin chain systems, which are characterized by the Poisson to GinUE transition within RMT.

The organisation of this paper is as follows. 
This introduction section is followed by Sec.~\ref{CSR} wherein the complex spacing ratios and related concepts have been reviewed. We define and examine the interpolating random matrix models in Sec.~\ref{RMTmodels}, followed by a discussion of investigated spin Hamiltonians in Sec.~\ref{NH-Hams}. Section~\ref{NH-results} contains the simulation outcomes for the spin-chain Hamiltonians and their comparison with RMT results. We conclude with a summary of our key observations along with possible future directions in Sec.~\ref{NH-sum}. 

%%%%%%%%%%%%%%%%%%%%%%%%%%%%%%%%%%%%%%%%%%%%%%%%%%%%%%%%%%
\section{Complex spacing ratios}
\label{CSR}

The distribution of complex spacing ratios, introduced in Ref.~\cite{SRP2020}, has come up as a reliable measure for distinguishing the integrable-versus-chaotic fluctuation properties of complex spectra and has found abundant use in recent works of non-hermitian physics~\cite{SRP2020,GSV2022,SG2022,GGK2022,DW2022}. It may be defined for both real and complex spectra unlike other popular ratio distributions like the traditional spacing ratios~\cite{OH2007,ABGR2013,ABGVV2013,SKK2020} and the nearest-neighbor (NN) or next-nearest-neighbor (NNN) ratios~\cite{SLTB2018} which are useful only in the case of real spectra. We briefly recapitulate the techniques involved in finding CSR for completeness of this paper.

Let $\{x_{1},x_{2},\cdots,x_{N}\}$ denote real or complex eigenvalues. For each eigenvalue $x$, let $x^\mathrm{NN}$ denote its nearest neighbor and $x^\mathrm{NNN}$ its next-nearest neighbour, which are identified on the basis of distances in real or complex plane. The CSR are then defined as,
\begin{align}
    z_{k} = \frac{x_{k}^\mathrm{NN}-x_{k}}{x_{k}^\mathrm{NNN}-x_{k}};~~~k=1,...,N.
\end{align}
When the spectrum is real, $z\equiv r$ with $-1\leq r \leq 1$. On the other hand for complex spectra, $z = r e^{i \theta}$, where $0 \leq r \leq 1$ and $-\pi\le \theta<\pi$. Besides the probability density function $\rho(r,\theta)$ for the CSR, it is useful to consider the radial $P_r(r) = \int_{-\pi}^{\pi} d \theta\, r \rho (r, \theta)$ and the angular $P_\theta(\theta) = \int_{0}^{1} dr\, r \rho (r, \theta) $ marginal density functions. Furthermore, it is instructive to look at the averages associated with these quantities, viz., $\langle r \rangle$ and $\langle \cos\theta \rangle$. The average quantity $\langle r \rangle $ and $\langle \tilde{r} \rangle$ has been frequently used along with the distribution of normal spacing ratios in several crossover-related studies ~\cite{SKK2020,BTS2018,GMVA2022,WR2023,MTHSLK2022,CK2013,CDK2014,KR2017}. These single number signatures are useful, for example, to estimate the parameter value at which a symmetry crossover takes place from one symmetry class to other in a physical system. 

For uncorrelated levels in the complex plane, associated with (2D) Poissonian spectral fluctuations, a flat distribution in the unit circle is observed for the CSR such that $\rho(z) = (1/\pi) \Theta(1- |z|)$. Consequently, in this case, the radial and angular distributions are given by $P_r(r) = 2 r \Theta(1-r)$, and $P_\theta(\theta) = 1/2 \pi$, respectively. These then lead to $\langle r\rangle=2/3$ and $\langle \cos \theta \rangle = 0$. On the contrary, quantum chaotic behaviour is inferred from an overlap with the Ginibre unitary statistics of RMT, featured by cubic level repulsion, $\rho_\mathrm{GinUE} \propto r^{3}$ as $r\to 0$, along with a dip in the ratio density at the centre and at small angles. In Ref.~\cite{SRP2020}, among other things, the authors have obtained the analytical expression for CSR density in GinUE case in terms of an $(N-1)$-fold integral. They found that Wigner-like surmise obtained from the $N=3$ case are not universal in nature and does not approximate large-$N$ behaviour due to boundary effects. To resolve this issue, they introduced the Toric Unitary Ensemble (TUE) which can be viewed as a two-dimensional analogue of the circular ensemble. For TUE also, they obtained an $(N-1)$-fold expression for the CSR density and demonstrated that small $N$ cases of this ensemble are universal and capable of approximating the large-$N$ behavior of GinUE. For our analyses, along with large GinUE-matrices simulation data, we use their $N=5$ result, which is denoted using TUE$_5$. The averages obtained using TUE$_5$ are $\langle r\rangle = 0.7315$ and $\langle\cos\theta\rangle=-0.1938$. The same averages evaluated using large size $(\sim 10^4)$ GinUE matrices are $\langle r\rangle = 0.7381$ and $\langle\cos\theta\rangle=-0.2405$. As noted in Ref.~\cite{SRP2020}, the convergence of $\langle\cos\theta\rangle$ computed from the TUE surmises is much slower than that of $\langle r\rangle$. We should also remark that very recently series-expansion-based approximate formulas have been derived for the GinUE CSR distribution in Ref.~\cite{DW2022}. 

It is known that the non-Hermitian Ginibre ensembles exhibit cubic level repulsion irrespective of the symmetry class, i.e. orthogonal or unitary, or even symplectic~\cite{GH1989,AKMP2019,HKKU2020}. This is unlike their Hermitian counterparts, where very distinct level repulsion behavior is observed depending on the symmetry class. As far as GinOE is concerned, it is known that it possesses real eigenvalues along with complex-conjugate pairs, and that the expected fraction of real eigenvalues for large-dimensions is given by $\sqrt{2/(N\pi)}$~\cite{EKS1994}. These eigenvalues have noticeable effects on the CSR, resulting in heightened values in the radial and angular marginal densities near $r=1$ and $\theta=-\pi$ (or $\pi$), respectively, compared to the GinUE results. However, as $N$ is increased, these deviations tend to diminish. Moreover, in a GinOE-GinUE crossover model, even for a tiny perturbation from GinOE towards GinUE limit gets rid of these deviations.  This difficulty is compounded, particularly in datasets derived from physical models, where substantial statistical noise may be present. As a result, attempting to investigate a GinOE-to-GinUE crossover using CSR is not viable.

%%%%%%%%%%%%%%%%%%%%%%%%%%%%%%%%%%%%%%%%%%%%%%%%%%%%%%%%%
\section{Phenomenological matrix models for Poisson-GinUE and GinOE-GinUE crossovers}
\label{RMTmodels}

For a system undergoing a transition from one symmetry class to another, one often examines the associated effects on the spectral fluctuations, which are commonly assessed by the distribution of nearest neighbour spacings or their ratios. In this context, for Hermitian systems, there already exist known interpolating formulas (either phenomenological or derived from small-dimension random matrix models) for the distribution of spacings between consecutive eigenvalues, as well as the associated ratios. These formulas cover various cases such as Poisson-GOE crossover, Poisson to semi-Poisson crossover, and the transitions among GOE, GUE, and GSE.~\cite{Mehta2004,CK2013,CDK2014,KR2017,SKK2020}. On the other hand, transitional models or crossover ensembles have remained far less explored in non-Hermitian systems and non-Hermitian RMT. A few examples can be found in Refs.~\cite{FKS1997,BE2021,ADM2022,BF2022}. In this section we consider phenomenological interpolating random matrix ensembles based on two cases of the following matrix model,
\begin{align}
\mathcal{H}_{\alpha} = \frac{\mathcal{H}_{0}+ \alpha \mathcal{V}}{\sqrt{1+\alpha^{2}}}.
\end{align}
Here $\mathcal{H}_{0}$ is an ``initial" random matrix for the crossovers detailed below. The ``final" $N$-dimensional random matrix $\mathcal{V}$ is taken from GinUE, i.e. its matrix elements are complex, independent and identically distributed (iid) Gaussian variables with zero-mean and unit-variance for both real and imaginary parts. The parameter $\alpha$ facilitates the crossover with  $\alpha = 0$ yielding the initial ensemble and $\alpha\rightarrow \infty$ leading to GinUE.
For the matrix $\mathcal{H}_{0}$, we consider the following two choices, corresponding to which we refer the above crossover model as matrix model 1 and 2 (MM1 and MM2):\\
\underline{MM1}: $\mathcal{H}_{0}$ is chosen as an $N$-dimensional diagonal matrix with iid zero-mean and unit-variance real Gaussians,\\
\underline{MM2}: $\mathcal{H}_{0}$ is chosen as an $N$-dimensional diagonal matrix with iid complex Gaussians zero-mean and unit-variance for both real and imaginary parts.

\begin{figure}[!tbp]
        \centering
\includegraphics[width=1.0\linewidth]{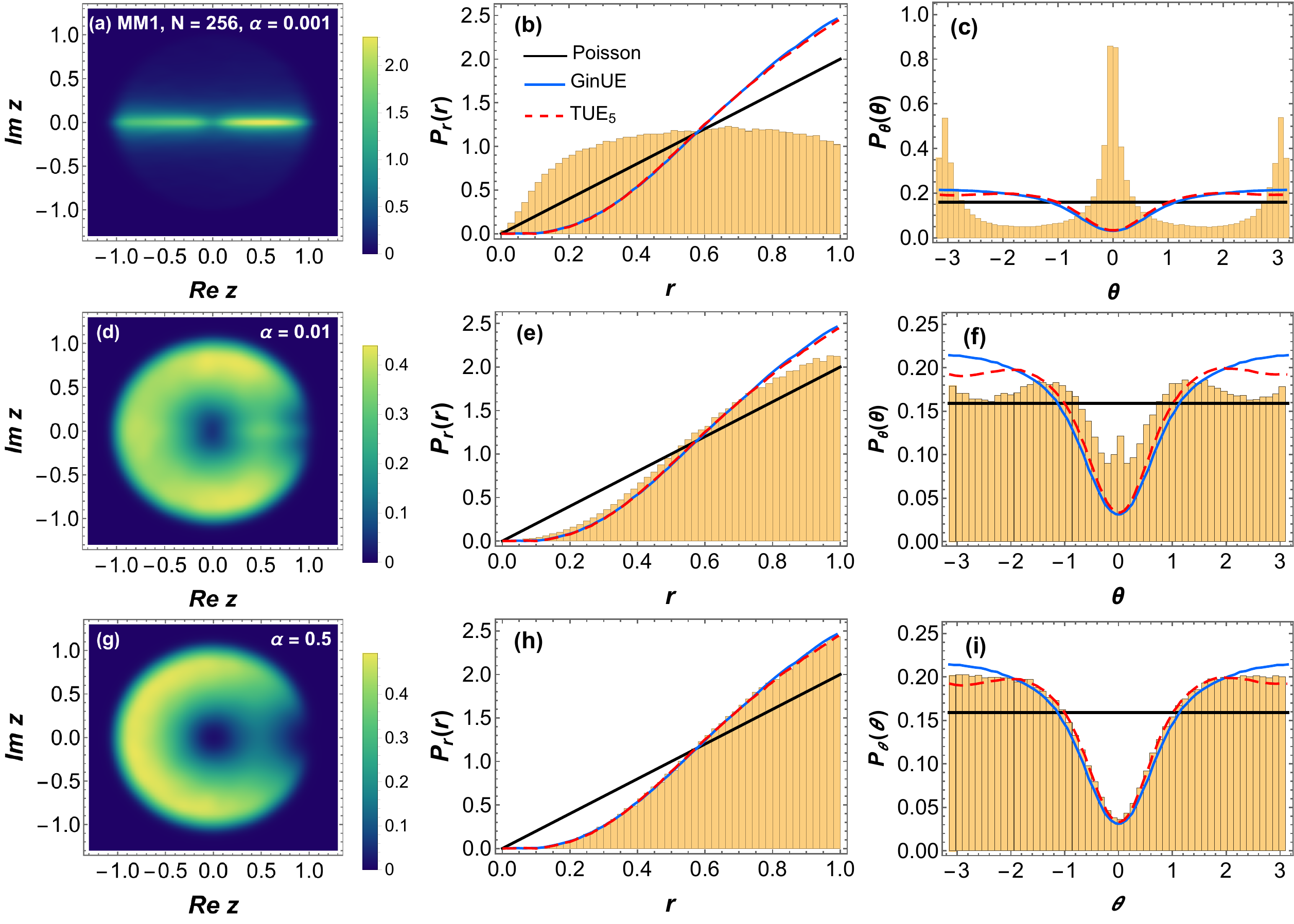}
        \caption{Plots of CSR density in the complex plane along with radial and angular marginals for 1D-Poissonian to GinUE crossover in MM1 for 1500 matrices of size $N = 256$.}
        \label{PR-GinUE-MM1-N256}
\end{figure}

\begin{figure}[!tbp]
        \centering
\includegraphics[width=1.0\linewidth]{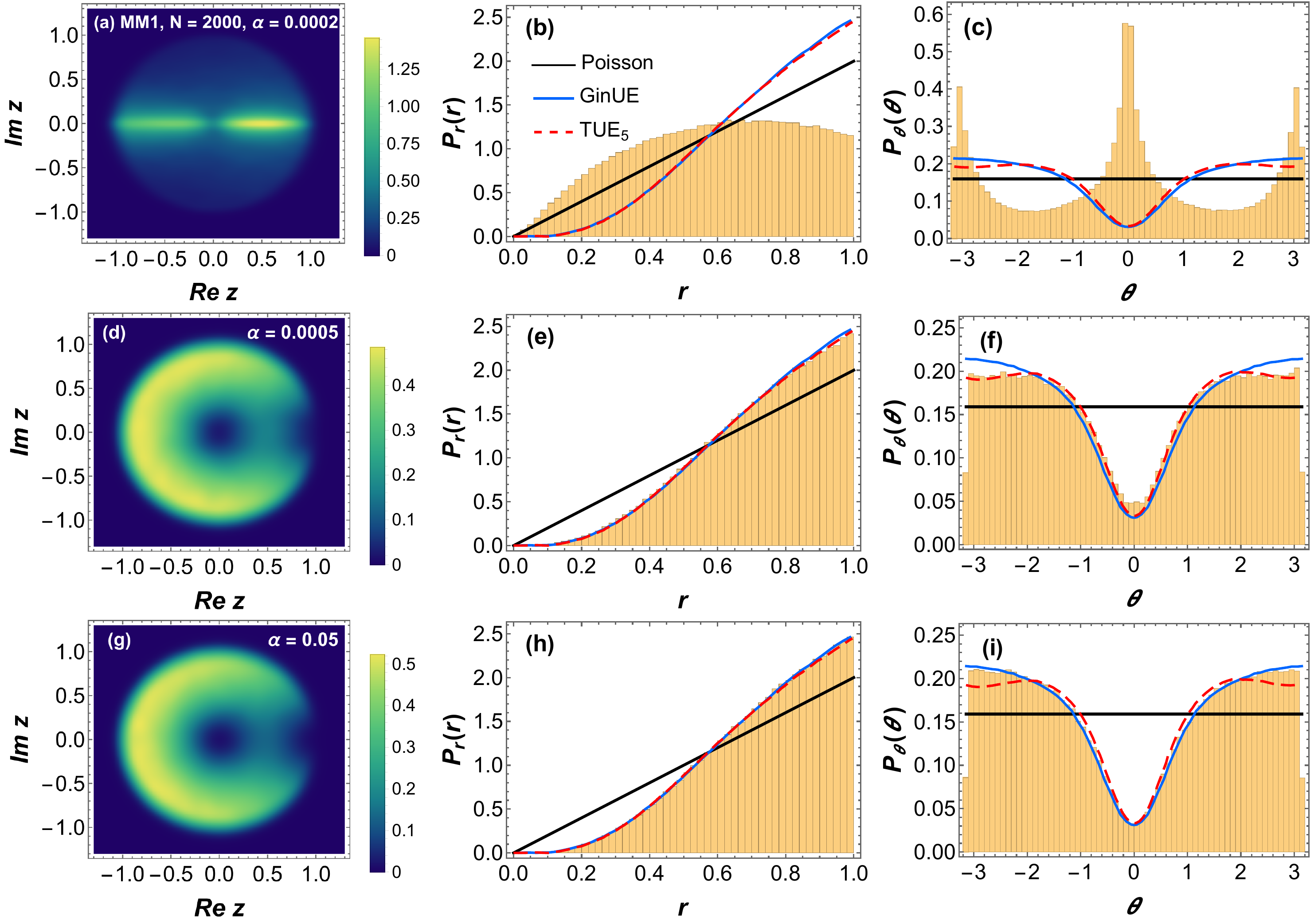}
        \caption{Plots for MM1, similar to Fig.~\ref{PR-GinUE-MM1-N256}, for an ensemble comprising 250 matrices of dimension $N = 2000$.}
        \label{PR-GinUE-MM1-N2000}
\end{figure}

We study the CSR density and the associated radial and angular marginals for the above crossover matrix models by varying the interpolation parameter $\alpha$. For each matrix model, we show the results for $N = 256$ and 2000  with ensemble comprising 1500 and 250 matrices, respectively. For MM1, which governs 1D-Poisson to GinUE transition, the results are presented in Figs.~\ref{PR-GinUE-MM1-N256} and~\ref{PR-GinUE-MM1-N2000}. The CSR density has been shown in the complex plane in the first columns, and the marginal densities for radial and angular distributions obtained from the simulation are depicted using histogram in the second and third columns, respectively. The black solid and red dashed lines shown with these histograms correspond to Poisson and TUE$_5$ analytical CSR results, respectively.  Additionally, we show blue solid lines which are based on the simulation of large ($\sim 10^4$) GinUE matrices which, in the case of angular distribution, differ appreciably from TUE$_5$ result towards $\theta=\pm \pi$. The stark differences between the top rows (a-c) and the others in these two figures is due to the eigenvalues having very small imaginary parts in the former case. This is the reason behind the accumulation of CSR in the vicinity of the real line noted for $\alpha = 0.001$ in the $N=256$ case and $\alpha=0.0002$ in the $N=2000$ case. With increasing $\alpha$, the eigenvalues gradually spread in the unit disc and approaches GinUE statistics. 

\begin{figure}[!tbp]
    \centering
        \includegraphics[width=1.0\linewidth]{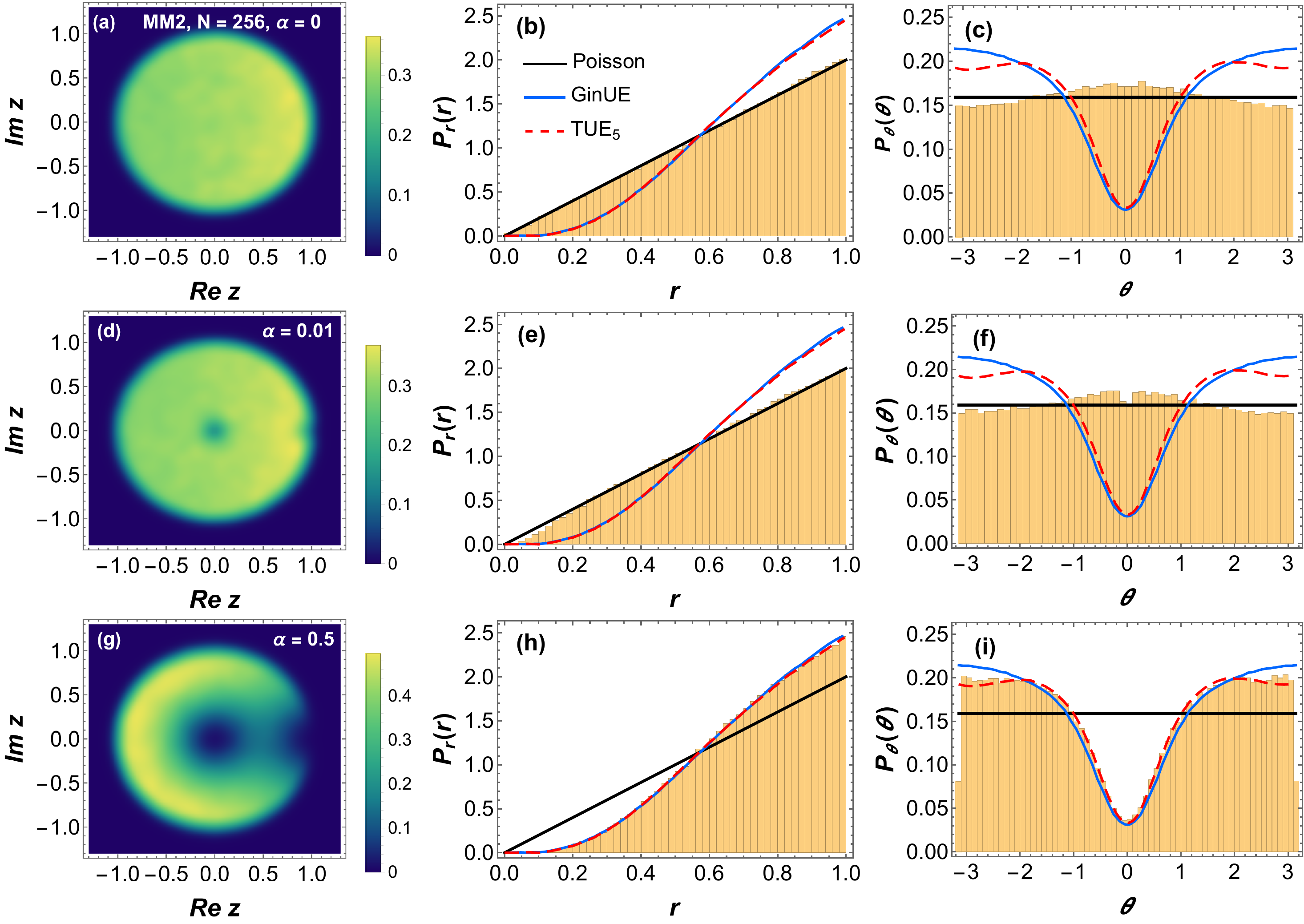}
        \caption{Plots of CSR density in the complex plane along with radial and angular marginals for 2D-Poissonian to GinUE crossover as modeled in MM2 for an ensemble comprising 1500 matrices of size $N = 256$.}
        \label{PC-GinUE-MM2-N256}
 \end{figure}

\begin{figure}[!tbp]
    \centering
\includegraphics[width=1.0\linewidth]{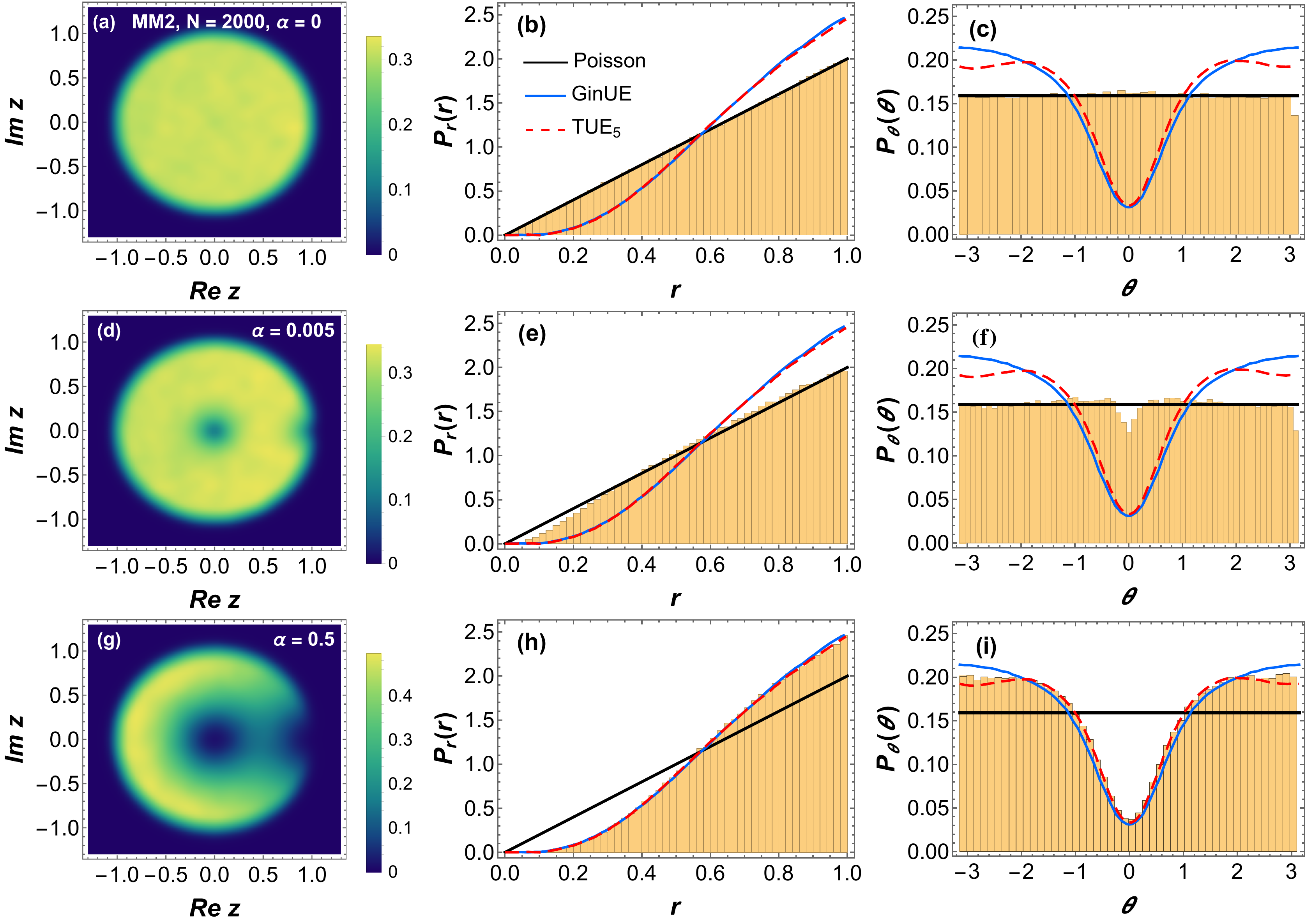}
        \caption{Plots for MM2, similar to Fig.~\ref{PC-GinUE-MM2-N256}, for an ensemble of 250 matrices of dimension $N = 2000$.}
        \label{PC-GinUE-MM2-N2000}
\end{figure}

For MM2, which facilitates 2D-Poisson to GinUE crossover, we show similarly the plots in Figs.~\ref{PC-GinUE-MM2-N256} and~\ref{PC-GinUE-MM2-N2000}. The crossover is observed in the CSR density as well as the marginals in these figures as $\alpha$ is increased beyond zero. Moreover, it is instructive to examine the $\alpha = 0$ limit. We find that in Fig.~\ref{PC-GinUE-MM2-N256}, for relatively smaller matrix dimension, $N =256$, the radial distribution $P_r(r)$ matches closely with the expected Poissonian analytical result. However, the angular distribution shows elevated density near $\theta = 0$ and deviate from the expected uniform behavior. Increasing the matrix dimension alleviates this deviation, as can be seen in Fig.~\ref{PC-GinUE-MM2-N2000}, where we have considered $N = 2000$.
%%%%%%%%%%%%%%%%%%%%%%%%%%%%%%%%%%%%%%%%%%%%%%%%%%%%%%%%%%%%%
%%%%%%%%%%%%%%%%%%%%%%%%%%%%%%%%%%%%%%%%%%%%%%%%%%%%%
%%%%%%%%%%%%%%%%%%%%%%%%%%%%%%%%%%%%%%%%%%%%%%%%%%%%%%%%%%%%%%

\section{Investigated Hamiltonian models}
\label{NH-Hams}
The XY spin-chain system in the presence of a transverse magnetic field (field along $z$-direction) may be represented as,
\begin{equation}
\label{HXY}
H^{(XY)} = \sum\limits_{j = 1}^{L}\left(J_{x} \sigma_{j}^{(x)}\sigma_{j+1}^{(x)} + J_{y} \sigma_{j}^{(y)}\sigma_{j+1}^{(y)} + \lambda \sigma_{j}^{(z)}\right),
\end{equation}
where $J_{x}, J_{y}\in \mathds{R}$ are the coupling constants while $\lambda\in \mathds{R}$ is the strength of the magnetic field along the $z$-direction. The operators $\sigma^{(x,y,z)}$ act on a $\left(\mathds{C}^{2}\right)^{\otimes L}$ dimensional Hilbert space, $L$ being the chain length. In terms of Pauli matrices ($\upsigma^{(x,y,z)}$) and identity matrices ($\mathds{I}$) these are given by, $\sigma_{j}^{(x,y,z)} = \mathds{I} \otimes \mathds{I} \otimes ....\otimes \upsigma^{(x,y,z)} \otimes...\otimes \mathds{I} \otimes \mathds{I}$. The Pauli matrix part, therefore, acts on the $j$th site of the chain while the identity operators act on the rest of the sites. Periodic boundary condition is imposed so that $\sigma_{N+1}^{(x,y,z)} = \sigma_{1}^{(x,y,z)}$.
The coupling constants for the Heisenberg terms may be modified to render them complex, leading to a non-Hermitian variant of the above Hamiltonian, as given below~\cite{ZS2013},
\begin{equation}
\label{RTH1}
    H_{1} = \sum\limits_{j = 1}^{L}\left(\frac{1+ i \gamma}{2} \sigma_{j}^{(x)}\sigma_{j+1}^{(x)} + \frac{1- i \gamma}{2} \sigma_{j}^{(y)}\sigma_{j+1}^{(y)} + \lambda \sigma_{j}^{(z)}\right).
\end{equation}
The parameter $\gamma\in \mathds{R}$ in this model controls the extent of non-Hermiticity. It is to be noted that while $[\mathcal{R},H_1]\ne0$ and $[\mathcal{T},H_1]\ne 0$, we have $[\mathcal{RT},H_{1}] = 0$. However, due to the antilinear nature of $\mathcal{T}$ and hence of the $\mathcal{RT}$ operator, the last relation does not guarantee existence of simultaneous eigenstates of both $H_1$ and $\mathcal{RT}$~\cite{Bender2007,Benderbook2019}. In case every eigenstate of $H_1$ does happen to be an eigenstate of the $\mathcal{RT}$ operator, it is said that the $\mathcal{RT}$ symmetry of $H_1$ is unbroken. On the other hand, if some of the eigenstates of $H_1$ are not simultaneously eigenstates of the $\mathcal{RT}$ operator, one concludes that the $\mathcal{RT}$-symmetry of $H_1$ is broken. To remind ourselves, the operator $\mathcal{R}$ is the linear rotation operator and has the effect of rotating each spin by an angle of $\pi/2$ about the $z$-axis,
\begin{equation}
    \mathcal{R} \equiv \exp\left[-\frac{i \pi}{4} \sum\limits_{j=1}^{L}\sigma_{j}^{(z)}\right]=\prod_{j=1}^L \frac{1}{\sqrt{2}}\left(\mathds{I}-i\sigma^{(z)}\right)_j.
\end{equation}
This amounts to an action, $(\sigma_{j}^{(x)},\sigma_{j}^{(y)},\sigma_{j}^{(z)}) \rightarrow (\sigma_{j}^{(y)},-\sigma_{j}^{(x)},\sigma_{j}^{(z)})$. On the other hand, the antilinear time-reversal operator $\mathcal{T}$ has the action $\mathcal{T} i\mathcal{T} = -i$. A hamiltonian symmetric with respect to an anti-linear operator such as $\mathcal{T}$, shows eigenvalues which are either real or appear in complex conjugate pairs~\cite{Benderbook2019,Fring2013}. When acting on the Pauli matrices, $\mathcal{T}$ acts as a complex conjugation such that $(\sigma_{j}^{(x)},\sigma_{j}^{(y)},\sigma_{j}^{(z)}) \rightarrow (\sigma_{j}^{(x)}, -\sigma_{j}^{(y)},\sigma_{j}^{(z)})$. 
 
 In Ref.~\cite{ZS2013}, the authors have exactly solved this spin-chain model using the techniques of Jordan-Wigner, Fourier and Boguliobov transformations extended to complex versions. Based on the exact results, exceptional points separating broken-unbroken regions of $\mathcal{RT}$ symmetry have also been identified. The spectra of the Hamiltonian in Eq.~\eqref{RTH1} possesses, respectively, real and complex eigenvalues corresponding to broken and unbroken $\mathcal{RT}$-symmetric phases, which in turn are decided by the parameter $\gamma$. Therefore, in the above model $\mathcal{RT}$ symmetry plays the same role as $\mathcal{PT}$ symmetry does in the generally studied $\mathcal{PT}$-symmetric pseudo-Hermitian systems showing real spectrum~\cite{BB1998,BBM1999}.

Two limiting cases of this hamiltonian are of particular importance to emphasize the above points. 
Firstly, when $\gamma= 0$, the above hamiltonian reduces to the ordinary XY model with an external magnetic field in the $z$ direction, the $J_{x} = J_{y} = 1/2$ case of Eq.~\eqref{HXY},
\begin{align}
\label{RTH0}
    H_{0} = \frac{1}{2}\sum\limits_{j = 1}^{L}\left( \sigma_{j}^{(x)}\sigma_{j+1}^{(x)} + \sigma_{j}^{(y)}\sigma_{j+1}^{(y)} + 2 \lambda \sigma_{j}^{(z)}\right).
\end{align}
$H_{0}$ is left unchanged under the action of $\mathcal{R}$ and $\mathcal{T}$ separately as well us under their joint action. This is unlike $H_1$ which remains invariant only under the joint action of $\mathcal{R}$ and $\mathcal{T}$. The Hamiltonian $H_0$ has a full real spectrum and all its eigenstates are shared by the $\mathcal{RT}$ operator. In the other limiting case of $H_{1}$ when $\gamma\gg\lambda$ and 1, $H_{1}$ reduces to 
\begin{align}
    H_\text{im} = \frac{i\gamma}{2} \sum\limits_{j = 1}^{L}\left( \sigma_{j}^{(x)}\sigma_{j+1}^{(x)} -  \sigma_{j}^{(y)}\sigma_{j+1}^{(y)} \right),
\end{align}
which displays a fully imaginary spectrum. Any eigenstate of $H_\text{im}$ corresponding to a non-zero eigenvalue is not an eigenstate of the $\mathcal{RT}$ operator. 

We consider the following modification of $H_{1}$ by introducing an additional magnetic field along the $x$-direction,
\begin{align}
\nonumber
    H_{2} = \sum\limits_{j = 1}^{L}\Big(\frac{1+ i \gamma}{2} \sigma_{j}^{(x)}\sigma_{j+1}^{(x)} + \frac{1- i \gamma}{2} \sigma_{j}^{(y)}\sigma_{j+1}^{(y)}  \\
 + \lambda \sigma_{j}^{(z)}+  \lambda_{1}\sigma_{j}^{(x)}\Big).
\end{align}
The introduction of this perturbative longitudinal field makes $H_{1}$ non-integrable on proper tuning of the system parameters like $\gamma$, $\lambda$ and $\lambda_{1}$. It also breaks the $\mathcal{RT}$ invariance of the system. Especially, when the $x$-field is random and the $z$-field is varied manually we see particularly distinct signatures of integrability-breaking portrayed by a transition from Poisson to GinUE statistics and can be compared with the matrix models of section~\ref{RMTmodels}. Upon inspection, we also find that the general trend of the numerical results remain the same if instead of the field along $x$-direction, we introduce a field along the $y$-direction.

Another variant of the above model is one where the $z$-field is rendered imaginary, viz.,
\begin{align}
 \nonumber 
 H_{3} =\sum\limits_{j = 1}^{L}\Big(\frac{1+ i \gamma}{2} \sigma_{j}^{(x)}\sigma_{j+1}^{(x)} + \frac{1- i \gamma}{2} \sigma_{j}^{(y)}\sigma_{j+1}^{(y)} \\
 +  i \lambda \sigma_{j}^{(z)}   +\lambda_{1}\sigma_{j}^{(x)}\Big).
\end{align}
The significance of such imaginary fields has already been discussed in Section~\ref{Intro}. In the present context, this field makes the complex eigenvalues having a significant imaginary part even when $\gamma$ is small, since now the paramater $\lambda$ also contributes to complex eigenvalues. This resulting effect can be seen in the density of the CSR, which is non-vanishing over the whole unit disc even for small $\gamma$ values. In the two previous hamiltonians, $H_{1}$ and $H_{2}$, this was not the case since $\gamma$ was the only non-hermiticity inducing parameter. For this hamiltonian also, the transition from integrable to non-integrable behavior is expected to be captured by the matrix model 2 of section~\ref{RMTmodels}.

In the following section, we examine the spectral fluctuations of the three hamiltonians $H_{1}$, $H_{2}$, and $H_{3}$ by varying parameters controlling non-hermiticity and strengths of the magnetic fields. For each case, one out of the set of available parameters is changed manually, while the others are chosen to be zero-mean, unit-variance Gaussian random numbers ($\sim \mathcal{N}(0,1)$). As far as the random nature of magnetic field is concerned, in real systems, it could be due to external sources producing a resultant magnetic field which is spatially inhomogeneous, leading to spin-glass like random energy landscape~\cite{BY1986,MPV1987,GS1998}, or rapidly varying in time over the timescale of a typical magnetic measurement.  For example, in Refs.~\cite{JLH2014,JLH2015,SLR2016,MBJ2019,SOU2023} many-body spin systems have been realized using trapped ions with laser driven interactions. The site-dependent static disorder has been mimicked using Stark shifts by laser beams focused to individual ions~\cite{SLR2016,MBJ2019}. Moreover, by temporally modulating the ac-Stark shifts, employing an arbitrary-wave-form generator with a switching time much faster than all other relevant timescales, time-dependent on site-energies have been engineered~\cite{MBJ2019}.

Interestingly, in the context of RMT investigations, the local spectral fluctuations are independent of such details and can exhibit signatures of integrability or chaos irrespective of the source of randomness. In particular, since CSR is a measure for short range spectral correlations, universal RMT behaviour is typically expected. Long range correlations, on the other hand, often depend on system specific details and deviate from RMT statistics. Indeed, while the impact of distinct sources of randomness can be different on the system, by tuning the corresponding strengths one can achieve desired level of symmetry breaking which, in turn, can lead to chaos. As a consequence, quantum spin chains in external random magnetic field have been the subject of investigation in several studies focusing on the spectral fluctuations and the associated symmetry crossovers, especially in the case of Hermitian quantum-spin chains~\cite{RNM2004,MM2014,ARB2002,KD2004,SM2016,MMR2014,RN2016,LLA2015,ZPP2008,SMW2017a,SMW2017b,SMW2017c,KKSG2022,KKSG2023}.

%%%%%%%%%%%%%%%%%%%%%%%%%%%%%%%%%%%%%%%%%%%%%%%%%%%%%%
\section{Empirical results : Discussion}
\label{NH-results}

In this section, we discuss the results of short-range spectral fluctuation properties of the above discussed non-Hermitian spin chain models using the complex spacing ratios, in which quantum chaotic behaviour can be identified via its vanishing density at the origin and a suppression of the same at small angles. This suppression at small angles is seen in the angular marginal distribution $P_\theta(\theta)$ and the slightly undulatory nature of the radial marginal distribution $P_r(r)$ associated with the cubic level repulsion distinctive of the Ginibre ensembles. One of our key observations is the Poisson-to-GinUE like spectral transition for certain range of parameters in $H_2$  and $H_3$, which we study for chain lengths $L = 6$ and 8. Through our numerical experiments, we identify parameter values for which these spin systems show spectral fluctuations quite similar to a transition from Poisson to GinUE statistics. 

The plots presented in this section contain results of density of CSR in the complex plane, and the marginal densities $P_r(r)$ and 
$P_\theta(\theta)$ for the three Hamiltonians $H_{1}$, $H_{2}$, and $H_{3}$ generated by varying parameters controlling non-hermiticity and strengths of the magnetic fields. These numerical results are then compared with analytical results for Poisson and $N = 5$ TUE ($\mathrm{TUE}_{5}$) represented with black solid and red dashed lines in the plots. We also plot the results from the matrix model simulation of $N = 10^{4}$ GinUE matrices with a blue solid line. As mentioned earlier, one out of the set of system parameters is manually varied, while the others are taken from the Gaussian distribution $\sim \mathcal{N}(0,1)$. We consider the entire spectrum of eigenvalues for generating these results. The overall size of the Hamiltonian matrices corresponding to $L = 6$ and $L = 8$ sized spin chains are 64 and 256, respectively. Ensembles comprising 4000 and 1500 matrices are considered respectively for these two sizes for both the chain lengths and various statistical properties of the energy spectra are examined. For $H_{2}$ and $H_{3}$, Poisson-to-GinUE like transition is quite apparent. In fact, even moderate chain lengths, like $L = 6$, lead to results quite close to GinUE on properly adjusting the system parameters. Moreover, in general, a rich variety of spectral-fluctuations behaviour is observed for the three hamiltonians as system parameters are varied. Detailed discussions of the results for each of them can be found in the following subsections.

%%%%%%%%%%%%%%%%%%%%%%%%%%%%%%%%%%%%%%%%%%%%%%%%%%%%%%%%
\subsection{Plots for $H_{1}$}

\begin{figure*}[!tbp]
\centering
\includegraphics[width=15cm]{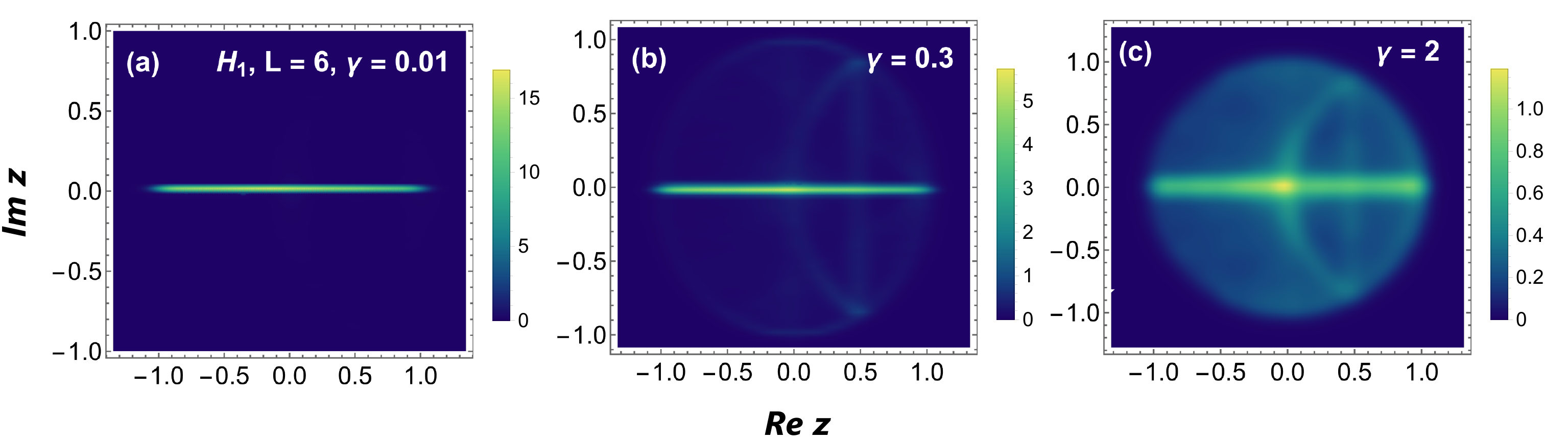}
\caption{Plots of CSR density in the complex plane for $L = 6$ of $H_{1}$. The parameter $\lambda$ is a random variate from the Gaussian distribution $\mathcal{N}(0,1)$,  while $\gamma$ is manually assigned the values 0.01, 0.3 and 2 in (a), (b) and (c). For lower values of $\gamma$ the densities are distributed on the real axis but they spread in the unit circle eventually as the value is increased and the eigenvalues acquire significant imaginary parts.}
\label{H1L1}
\end{figure*}

\begin{figure*}[!tbp]
\centering
\includegraphics[width=15cm]{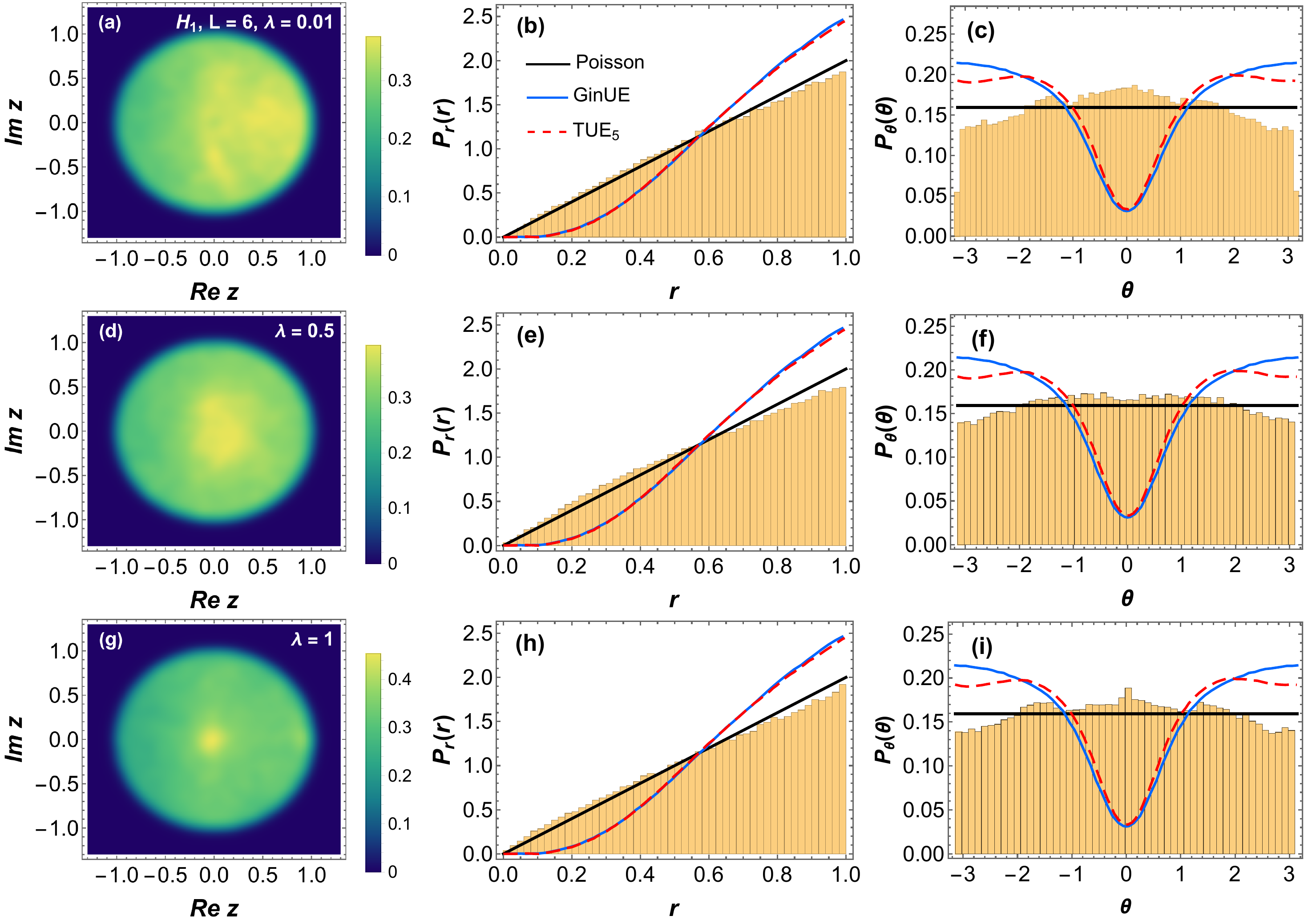}
\caption{Plots of CSR density and its marginals ($r$ and $\theta$ distributions) for $L = 6$ for $H_{1}$. $\gamma \sim \mathcal{N}(0,1)$, while $\lambda$ is 0.01 in (a)-(c), 0.5 in (d)-(f) and 1 for (g)-(i). Black solid lines depict the analytical results for the Poisson distribution, the red dashed one corresponds to $N =$5 results for the Toric Unitary Ensemble (TUE), and the blue solid line corresponds to results from $10^4$-sized GinUE matrices. In all cases close-to-Poisson results are observed.}
\label{H1L2}
\end{figure*}

Here we examine the spectral fluctuations of the Hamiltonian $H_{1}$. In Fig.~\Ref{H1L1} the CSR for $L = 6$ spin-chain length of this model is inspected with the variation of $\gamma$, while $\lambda$ is taken from the Gaussian distribution $\mathcal{N}(0,1)$. In Fig.~\Ref{H1L1}(a), at $\gamma = 0.01$, majority of the eigenvalues are real within the numerical precision considered and are of the form $\pm x$. Only a few eigenvalues have noticeable imaginary parts but even those are very small. As a consequence, the CSR density is nonzero only on and in the vicinity of the real line. In Fig.~\Ref{H1L1}(b), for $\gamma=0.3$, along with the nonzero density along the real line, a faint pattern appears inside the unit circle. This becomes more pronounced for 
$\gamma=2$ in Fig.~\Ref{H1L1}(c) where a bow-arrow like structure is observed. On inspection of the eigenvalues for this particular case, a number of interesting patterns in the eigenvalues are observed. Along with the real eigenvalues of the form $\pm x$, complex eigenvalues of the form $\pm (x\pm iy)$ are observed. These patterns in the eigenvalues affect the CSR distribution which involves doubly-degenerate real as well as complex-conjugate ratios. We should remark here that in some works related to the study of many-body localization transition in non-Hermitian models, this crossover from real to complex eigenvalues and the suppression of imaginary parts of complex eigenvalues for general non-Hermitian hamiltonians having time-reversal symmetry has been discussed as a signature of many-body localization~\cite{HKU2019}. In Fig.~\Ref{H1L2}, CSR density plots along with marginal densities 
$P_r(r)$ and $P_\theta(\theta)$ are plotted, but now with $\lambda$ manually varied and $\gamma$ taken as Gaussian random numbers from $\mathcal{N}(0,1)$. The density plots show some resemblance to Poisson-like statistics in all three cases of $\lambda$, viz., 0.01 in (a)-(c), 0.5 in (d)-(f) and 1 in (g)-(i). In the CSR plot in the complex plane, as shown in the panels (a), (d) and (g), there is an enhanced density around the origin, which becomes more concentrated and hence brighter as $\lambda$ increases. In Figs.~\Ref{H1L2} (c), (f) and (i), the $\mathrm{P}_\theta(\theta)$ plot shows uneven surface due to these regions. Compared to $\mathrm{P}_\theta(\theta)$, $\mathrm{P}_r(r)$ is closer to (2D) Poisson-statistics behavior, as is observed in the panels (b), (e), and (h). However, if $\lambda$ is increased further the results start showing deviation from the Poisson statistics. In the context of $\mathcal{RT}$ symmetry, for $\gamma$ value fixed around zero, $H_1$ is mostly in the $\mathcal{RT}$-preserved phase which results in most of the eigenvalues being real, as in the case of Fig.~\Ref{H1L1}. On the other hand, the variation in $\gamma$ caused due to sampling from Gaussian distribution, along with particular choices of $\lambda$, results in $\mathcal{RT}$-broken phase of $H_1$, and hence majority of the eigenvalues are complex which result in the plots observed in Fig.~\ref{H1L2}.

Besides the plots, in Table~\ref{tabH1L6}, we provide a list of single number signatures $\langle r \rangle $ and $-\langle \cos \theta \rangle $  for various values of $\lambda$ and $\gamma$. Along with the marginal distributions, these averages help in stipulating the nature of the distribution and how close it is to being Poisson or GinUE. For $H_{1}$, these averages when $\gamma$ is varied and $\lambda$ is Gaussian has to be analysed rather carefully because neither of the them are close to (2D) Poisson values for $\gamma = $0.01. However when $\gamma$ is increased to 0.3 and 2, $\langle r \rangle $ gets closer to 0.67, but even then $-\langle \cos \theta \rangle$ values are not that close to zero. Also on careful examination of the density of CSR plots, we see that their behaviour is in-fact closer to 1D-Poisson case for smaller value of $\gamma$ which when increased leads to the remarkably different bow-arrow like structure already discussed above. Also for $H_{1}$ we see that negative values of $-\langle \cos \theta \rangle$ arise in most cases, for variation of both $\gamma$ and $\lambda$. In fact, the only non-negative value for $H_{1}$, $L = 6$ arises for $\gamma = 0.01$. Such negative values of $-\langle \cos \theta \rangle$ have also appeared in Ref.~\cite{SRP2020} for spin models with bulk dephasing. For the variation of $\lambda$ when $\gamma$ is Gaussian, the spectral statistics somewhat close-to-Poisson, evident from the radial marginal density and $\langle r \rangle$. However the angular marginal density and $\langle \cos\theta\rangle$ still show deviation from the uniform distribution expected for the Poisson. However this distribution is expected to get flattened out as the Hamiltonian matrix dimension is increased.
%%%%%%%%%%%%%%%%%%%%%%%%%%%%%%%%%%%%%%%%%%%%%%%%%%%%
%%%%%%%%%%FIGURE%%%%%%%%%%
%%%%%%%%%%%%%%%%%%%%%%%%%%%%%%%%%%%%%%%%%%%%%%
\begin{table}[!tbp]
\begin{tabular}{ccc}
\hline
$\gamma$ & $\langle r \rangle$ & $-\langle \cos \theta \rangle$ \\
\hline
0.01 & 0.479  & 0.053  \\
0.3 & 0.600 & $-0.069$ \\
2 & 0.633 & $-0.059$ \\
\hline
 $\lambda$ & $\langle r \rangle$ & $-\langle \cos \theta \rangle$ \\
\hline
0.01 & 0.655  & $-0.072$ \\
0.5 & 0.647  & $-0.042$ \\
1 & 0.650 & $-0.048$ \\
\hline
\end{tabular}
\caption{\label{tabH1L6}%
Single number signatures of the $H_{1}, L = 6$ for parameter values corresponding to the plots.}
\end{table}
%%%%%%%%%%%%%%%%%%%%%%%%%%%%%%%%%%%%%%%%%%%%%%%%%%%
\subsection{Plots for $H_{2}$}

\begin{figure*}[!tbp]
\centering
\includegraphics[width=11cm]{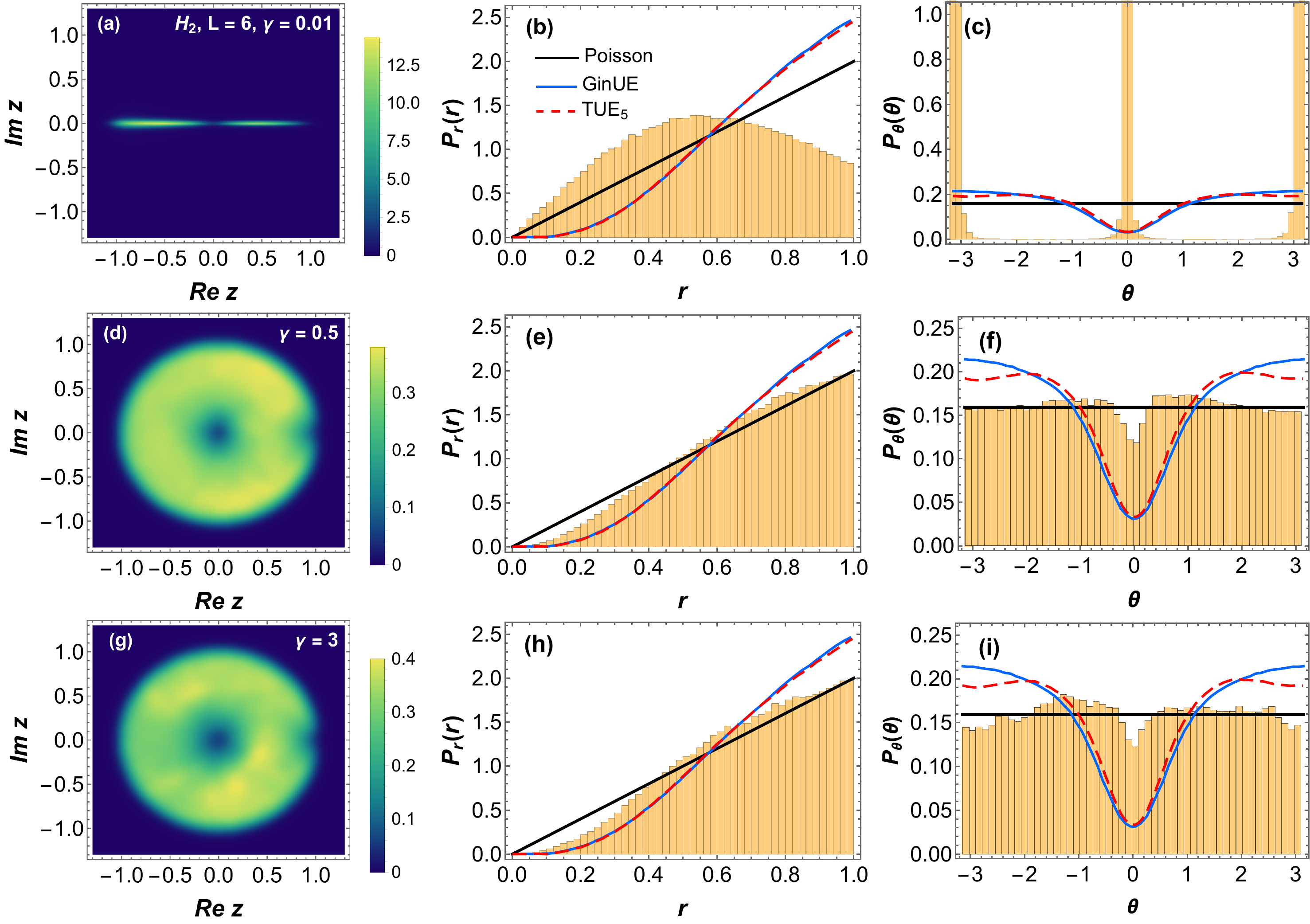}
\caption{Plots of CSR density and the associated marginals for $L = 6$ of $H_{2}$. The parameter $\gamma$ is varied manually while $\lambda$, $\lambda_{1}$ are random variates from the Gaussian distribution $ \mathcal{N}(0,1)$. We have $\gamma$ = 0.01 in (a)-(c)(top row), 0.5 in (d)-(f) (middle row) and 3 for (g)-(i) (bottom row). The plots (d)-(f) show subtle signatures of $\mathrm{TUE}_{5}$ denoted by the red line. The ratio density at the centre, vanishes to some extent in (d). This is also seen in the $P_\theta(\theta)$ vs $\theta$ plot in (f) which also shows slight suppression of small angles. For $\gamma = 3$ in (g) the results of density plots show areas of elevated brightness on the disc but the corresponding marginals show features very similar to the previous one.}
\label{H2-L6-gamvar}
\end{figure*}

\begin{figure*}[!tbp]
\centering
\includegraphics[width=11cm]{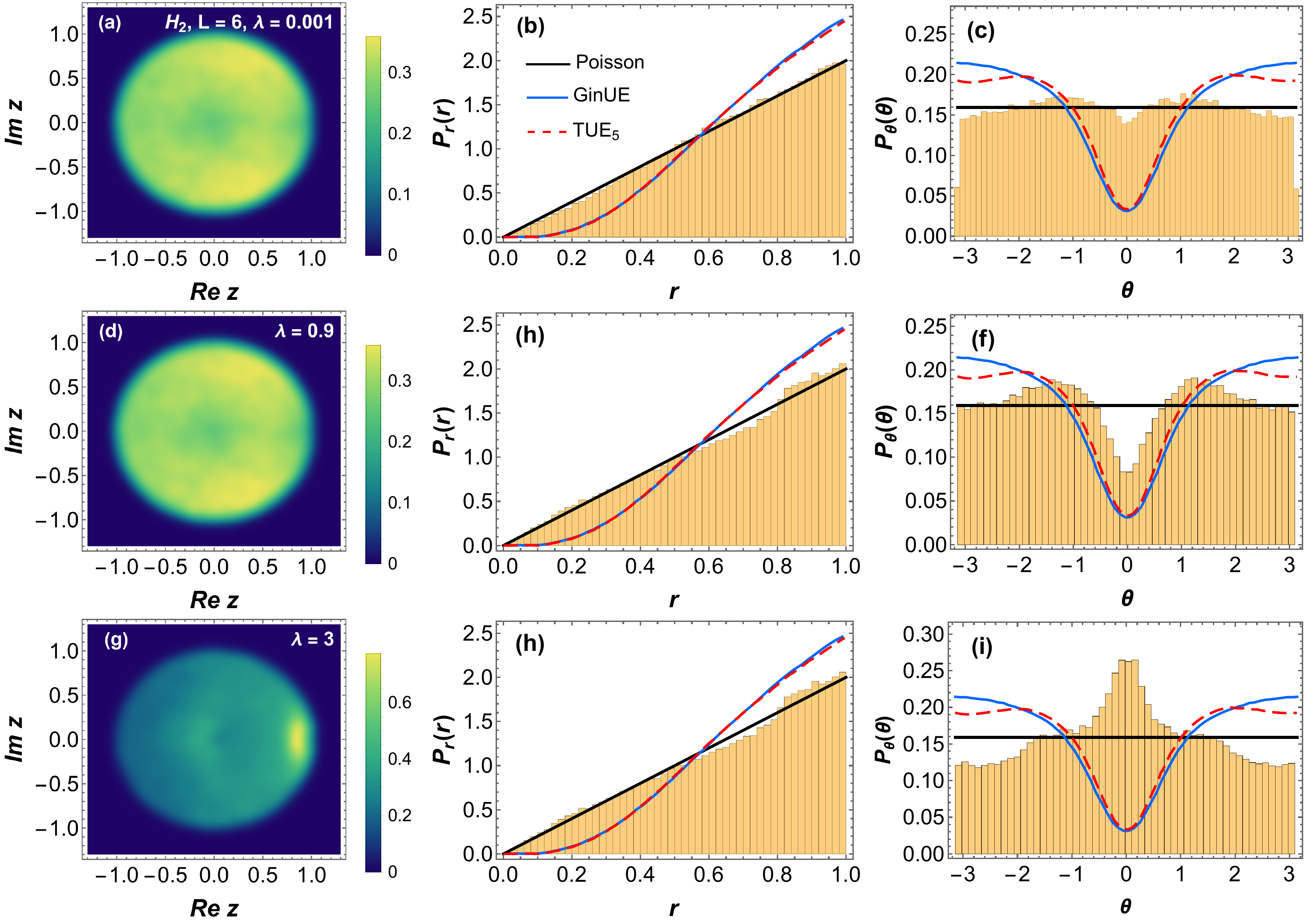}
\caption{Plots of CSR and its marginal densities for $H_{2}$ with the variation of $\lambda$, with $\gamma$ and $\lambda_{1} \in \mathcal{N}[0,1]$. Poisson-like behaviour featured by a flat distribution of the ratio density in the unit circle and uniform radial and angular distributions is observed for $\lambda = 0.001$, in (a)-(c) (top row). As the value of $\lambda$ is increased, large-N GinUE behaviour is observed for $\lambda = 0.9$ in (d)-(f) (middle row). $P_r(r)$ shows slight undulation, a distinguishing feature in the TUE case. $P_\theta(\theta)$ on other hand shows a strong suppression of small angles. On the contrary for $\lambda = 3$ in (g)-(i) (bottom row) accumulation of CSR at small angles is observed.}
\label{H2-L6-lambvar}
\end{figure*}

\begin{figure*}[!tbp]
\centering
\includegraphics[width=14cm]{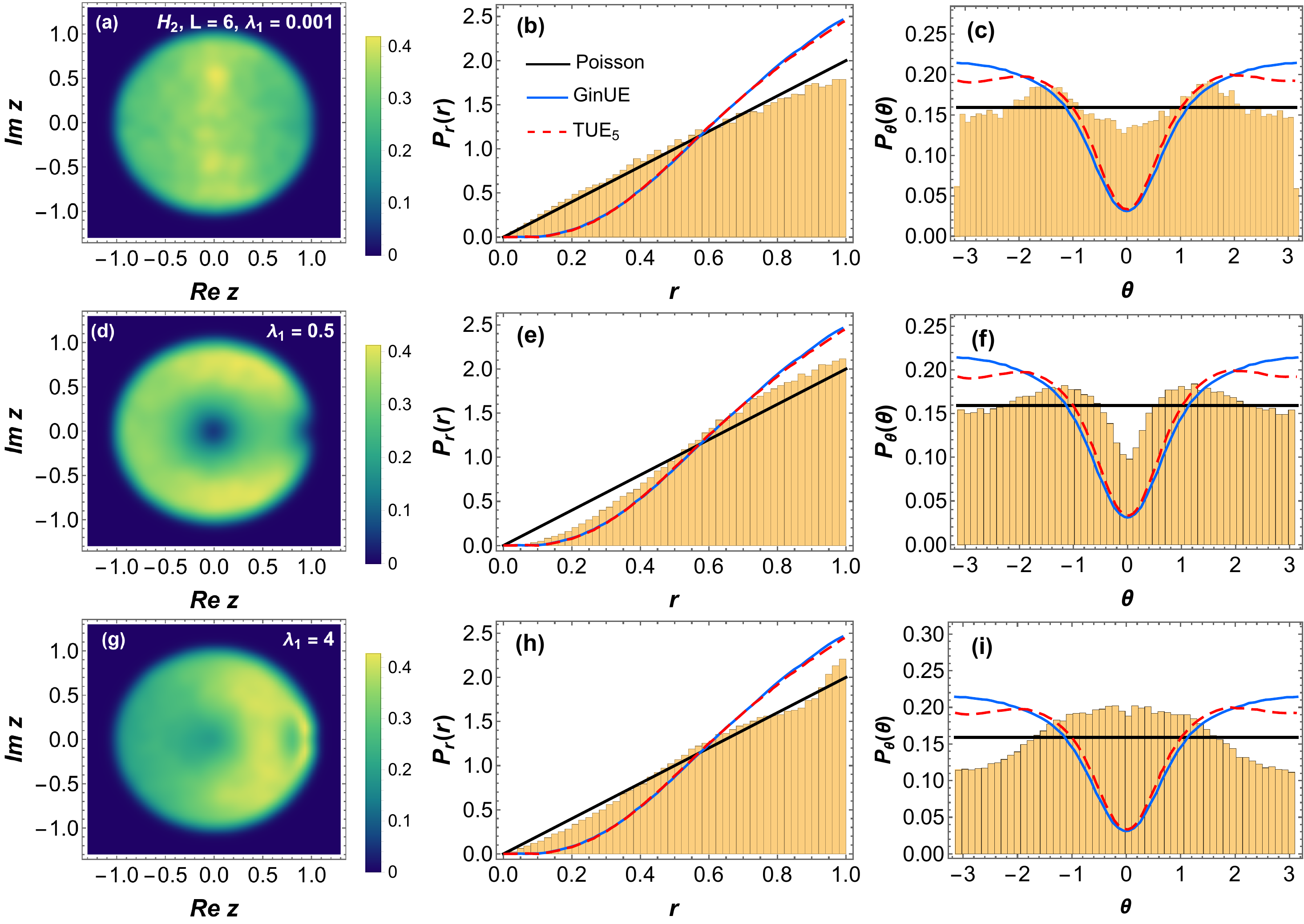}
\caption{Plots of CSR density and the associated marginals. In this case only $\lambda_{1}$ is varied while the remaining parameters are Gaussian variates from distribution $ \mathcal{N}[0,1]$. Although close-to-Poisson and TUE$_{5}$-like behaviour is observed for $\lambda_{1} = 0.001$ in (a)-(c) (top row) and 0.5 (d)-(f) (middle row) respectively, RMT behaviour in general is much less prominent when $\lambda_{1}$ is varied compared to $\lambda$ as in the previous figure.}
\label{H2-L6-lamb1var}
\end{figure*}

\begin{figure*}[!tbp]
\centering
\includegraphics[width=14cm]{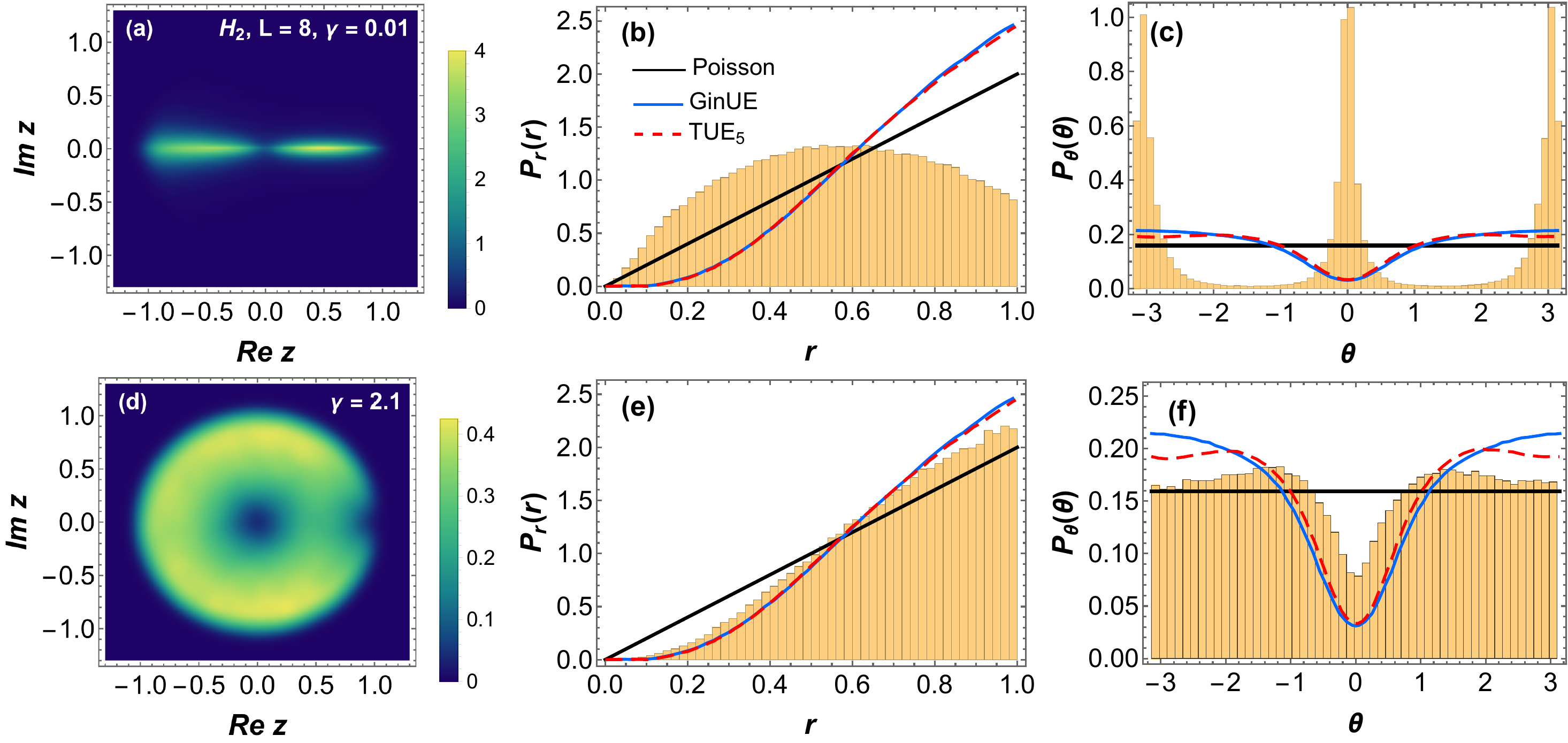}
\caption{Similar to Fig.~\ref{H2-L6-gamvar} but with $L = 8$ in this case. Due to larger chain size, the suppression of small angles is much more prominent and persists till broader range of values compared to $L = 6$ case. Especially for $\gamma = 2.1$ in (d)-(f), $\mathrm{TUE}_{5}$ features are very prominent.}
\label{H2-L8-gamvar}
\end{figure*}

\begin{figure*}[!tbp]
\centering
\includegraphics[width=15cm]{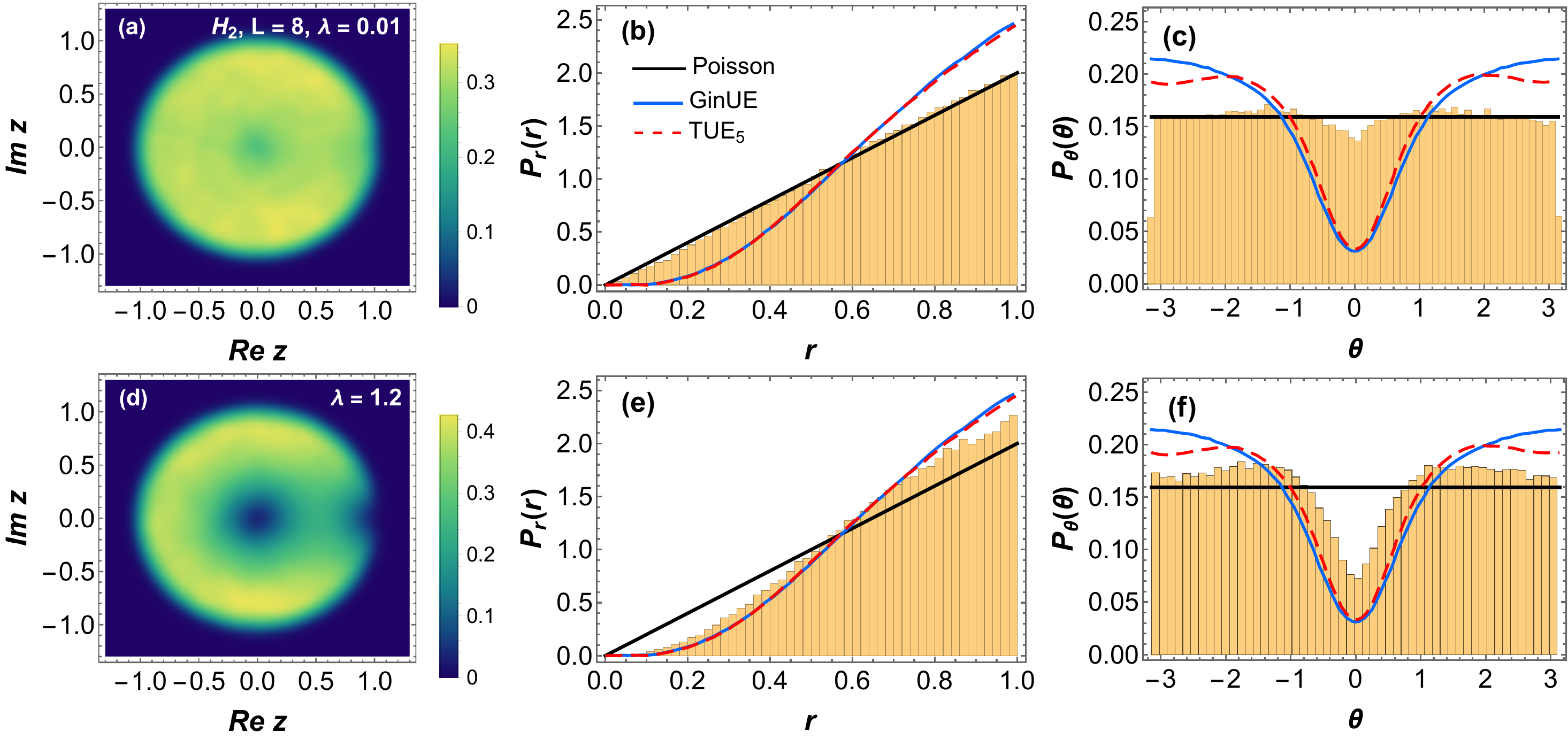}
\caption{Here $\lambda$ is varied for $L = 8$ of $H_{2}$. Poisson-like features are very prominent for 
$\lambda = 0.01$ in (a)-(c) in the top row and when $\lambda$ is increased to 1.2 in (d)-(f) (bottom row) the features are very close to the analytical results for $\mathrm{TUE}_{5}$.}
\label{H2-L8-lambvar}
\end{figure*}

\begin{figure*}[!tbp]
\centering
\includegraphics[width=15cm]{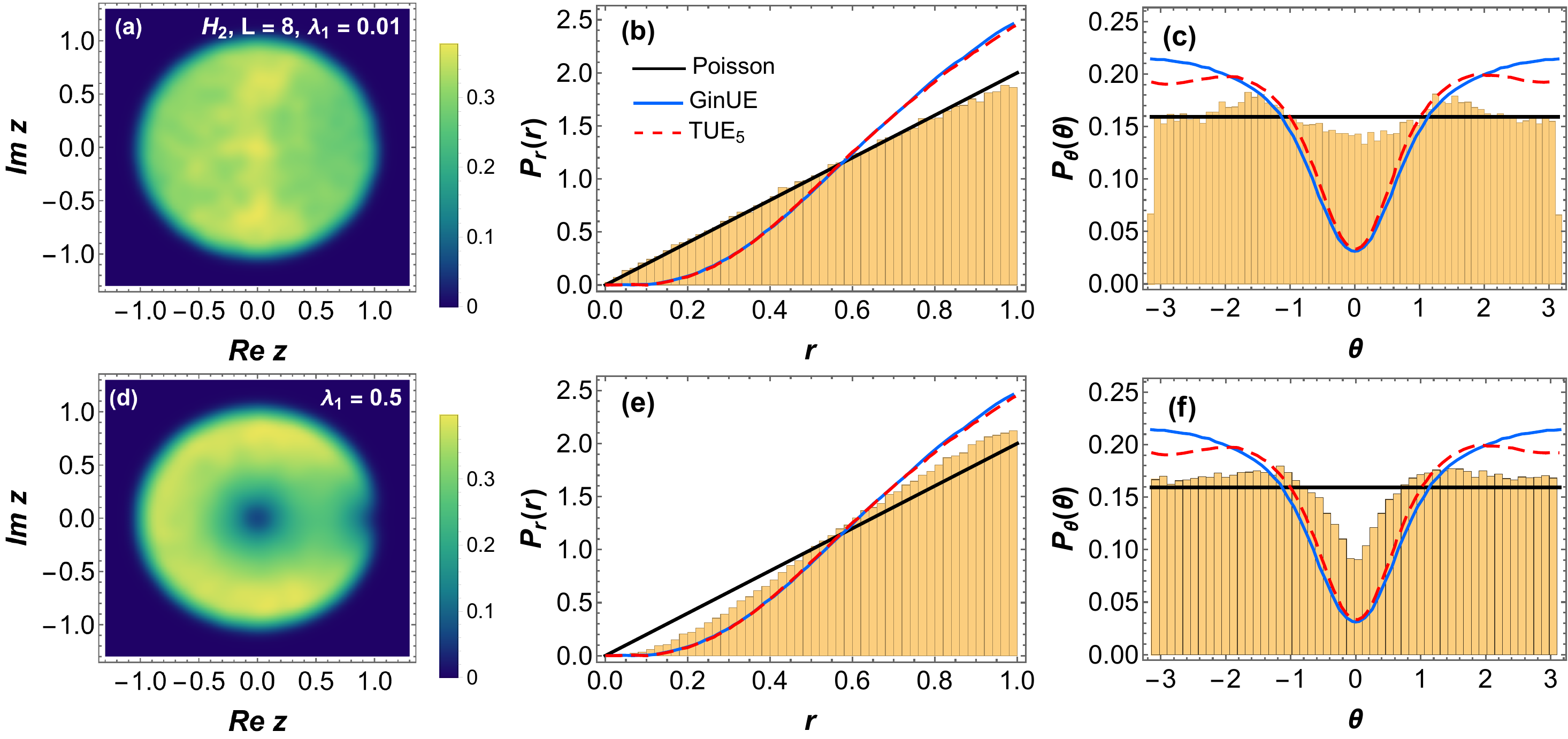}
\caption{In this case, $\lambda_{1}$ is varied for $L = 8$ of $H_{2}$. Larger chain size ensures more prominent RMT behavior for $\lambda_{1}$ = 0.5 ((d)-(f), bottom row) compared to that observed for $\lambda_{1} = 0.5$ in $L = 6$ spin chain size.}
\label{H2-L8-lamb1var}
\end{figure*}

We now examine the variation in spectral fluctuations for $H_{2}$ depending on system parameters $\gamma$, $\lambda$, and $\lambda_1$. These are shown in Figs.~\ref{H2-L6-gamvar},~\ref{H2-L6-lambvar}, \ref{H2-L6-lamb1var} for $L = 6$, and in Figs.~\Ref{H2-L8-gamvar}, \ref{H2-L8-lambvar}, \ref{H2-L8-lamb1var} for $L = 8$. In each case one of the parameter is manually varied, while the remaining two are taken from Gaussian distribution $\mathcal{N}(0,1)$. 

In Fig.~\Ref{H2-L6-gamvar}, CSR density in the complex plane and the marginals $P_r(r), P_\theta(\theta)$ are studied for three values of $\gamma$, viz., 0.01, 0.5 and 3. For $\gamma = 0.01$, like Fig.~\Ref{H1L1} (a), the spectra exhibits only a limited proportion of complex eigenvalues characterized by small imaginary parts which reflects in the ratio density spreading on the real line. This is more pronounced on the negative real axis and almost disappears at the origin. For $\gamma=0.5$ in Figs.~\Ref{H2-L6-gamvar}(d)-(f), the quantum chaotic behaviour in the density plot is somewhat implied from the vanishing density at the origin and at small angles. However, neither $P_r(r)$ nor $P_\theta(\theta)$ is close to analytical results. In Fig.~\Ref{H2-L6-gamvar}(f) slight dip at small angles is noticed. As the value of $\gamma$ is further increased, ratio density gets localized at random regions within the unit circle. The marginal densities $P_r(r)$ and $P_\theta(\theta)$ also show statistics quite distinct from both Poisson and GinUE.

In Fig.~\Ref{H2-L6-lambvar}, we vary $\lambda$ and examine its impact on the CSR. For $\lambda=0.001$, which corresponds to weak $z$-field, close-to-Poisson like statistics is observed for the CSR density in Fig.~\Ref{H2-L6-lambvar}\,(a) and a slight dip is noticed in $P_\theta(\theta)$ in Fig.~\Ref{H2-L6-lambvar}\,(c). However, $P_r(r)$ shown in Fig.~\Ref{H2-L6-lambvar}\,(b) matches Poisson statistics closely. The suppression of CSR density at small angles and origin as in Fig.~\Ref{H2-L6-lambvar}\,(d), results from an increase in $\lambda$ to 0.9 and causes clear transition from almost-Poisson to GinUE-like statistics. In this case, both radial and angular marginal densities in Figs.~\Ref{H2-L6-lambvar}\,(e) and ~\Ref{H2-L6-lambvar}\,(f) exhibit resemblance to GinUE results. As $\lambda$ is increased further, there is an increase in the density at smaller angles within the angular distribution, as depicted in Fig.~\Ref{H2-L6-lambvar}\,(i). This trend aligns with the accumulation of the CSR at small angles compared to the remaining parts of the unit disk, as can be seen in Fig.~\Ref{H2-L6-lambvar}\,(g). Intriguingly, this behavior stands in direct contrast to that observed in GinUE. However in general, for the case of $H_{2}$, it is possible to achieve the best overlap to RMT statistics by carefully tuning $\lambda$ while the other two parameters are chosen from a random distribution.

For $H_{2}$, changes in $\lambda_1$ in Fig.~\ref{H2-L6-lamb1var} does not show much overlap to GinUE results and vary quite significantly from RMT statistics except for the case of $\lambda = 0.5$, where suppression of small angles is somewhat noticeable from the density of CSR and the angular marginal distribution plot.

In Figs.~\Ref{H2-L8-gamvar},~\Ref{H2-L8-lambvar},~\Ref{H2-L8-lamb1var}, which portray the above three cases for spin chain length $L=8$, we find that for adequate values of the crossover parameters, it is possible to approach the exact results of GinUE more closely, due to the increased hamiltonian matrix dimensions. In Fig.~\Ref{H2-L8-gamvar}, for very small values of $\gamma$ (e.g., 0.01), the ratios again assemble near the real line but eventually spreads across the unit disk, as can be seen in the plot for $\gamma=2.1$. This feature is common in all cases of $\gamma$ variation which indicates that this parameter leads to a transition from real to complex eigenvalues. As previously discussed, this kind of transition from real to complex eigenvalues have been indicated in other non-Hermitian spin chains models and in our MM1 which captures interpolation between 1D-Poisson and GinUE statistics. Coming back to the present case, GinUE-like behavior can be seen for $\gamma=2.1$, whereas for $H_{2}, L = 6$, the GinUE-like features were not prominent. As $\gamma$ is increased further, CSR gets localized at certain regions of the unit circle, which is quite different from either Poisson or GinUE behaviour.  In Fig.~\Ref{H2-L8-lambvar}, where we examine the impact of variation of the parameter $\lambda$, Poisson-like characteristics are noticeably prominent for $\lambda=0.01$, although a suppression for small angles is seen in $P_\theta(\theta)$. For $\lambda=1.2$, GinUE correspondence is well evident. The plots in Fig.~\Ref{H2-L8-lamb1var}, based on the variation of $\lambda_1$, show trend similar to that in Fig.~\Ref{H2-L6-lamb1var}, however in the former case of $L=8$, GinUE-like behaviour is much stronger. Overall, we find that quantum chaotic features for $H_{2}$ are much more pronounced when the parameter $\lambda$ is varied while the others are chosen to be random numbers.

The results for $H_{2}$ can be well approximated by MM1, which interpolates between 1D-Poisson and the GinUE symmetry classes, when $\gamma$ is varied manually as discussed above. This can be seen by comparing the plots in Figs.~\ref{PR-GinUE-MM1-N256} and \ref{PR-GinUE-MM1-N2000} of the matrix model with the spin chain simulation in Figs.~\ref{H2-L6-gamvar} and \ref{H2-L8-gamvar}. When $\lambda$ and $\lambda_{1}$ are varied, the remaining plots of $H_{2}$ for both $L = 6$ and $L = 8$ show closer overlap to MM2 results in Figs.~\ref{PC-GinUE-MM2-N256} and \ref{PC-GinUE-MM2-N2000}, which capture 2D-Poisson to GinUE crossover.

The single number signatures for $L = 6$ and 8 of this chain are given in Tables~\ref{tabH2L6} and \ref{tabH2L8}. We see that for various values of $\gamma,\lambda,\lambda_{1}$, $\langle r \rangle$ is closest to the GinUE value of ($\sim 0.74$) for $L = 8$, Values as close as 0.704 ($\lambda_{1} = 0.5$), 0.714 ($\gamma = 2.1$) and 0.715 ($\lambda = 1.2$) is observed. Similar to $H_{1}$, negative values of for $-\langle \cos \theta \rangle $ also appears for $H_{2}, L = 6$, in the cases of $\lambda = 0.001, 3$ and $\lambda_{1} = 4$ these negative values also appear. However further exploration is needed to understand this behaviour in the context of our spin chains. In general, erratic behavior which is neither Poissonian nor GinUE is exhibited if any of the system parameters $\gamma,\lambda$ and $\lambda_1$ are increased over a certain value depending on the system size.

%%%%%%%%%%%%%%%%%%%%%%%%%%%%%%%%%%%%%%%%%%%%%%%%%%%%%%%
%%%%%%%%%%%%%%%%%%%%FIGURE%%%%%%%%%%%%%%%%%%%%%%%%%%%%%%%%%
%%%%%% FIGURE %%%%%%%%%%%%%%%%%%%%%%%%%%%%%
%%%%%%%FIGURE%%%%%%%%%%%%%%%%%%%%%%%%%%%%%%%%
%%%%%%%%%%%%%%%%%%%%Figure%%%%%%%%%%%%%%%%%%%%%%%
%%%%%%%%%Figure%%%%%%%%%%%%%%%%%%%%%%%%%%%%%%%
%%%%%%%%%%%%%%%%%%%%%%%%%%%%%%%%%%%%%%%%%%%%%%%%%%%%%%%%%%%%%%%%%%
\begin{table}[!tbp]
\begin{tabular}{ccc}
\hline
 $\gamma$ &  $\langle r \rangle$ & $-\langle \cos \theta \rangle$ \\
\hline
0.01 & 0.564  & 0.355 \\
0.5 & 0.689  & 0.002\\
3 & 0.662 & 0.094\\

\hline
 $\lambda$ & $\langle r \rangle$ & $-\langle \cos \theta \rangle$ \\
\hline
0.001 & 0.674 & $-$0.018  \\
0.9 & 0.713 &  0.040 \\
3 & 0.668 & $-$0.153 \\

\hline
 $\lambda_{1}$ & $\langle r \rangle$ & $-\langle \cos \theta \rangle$ \\
 \hline
 0.001 & 0.651 & 0.009 \\
0.5 & 0.706  & 0.011 \\
4 & 0.679 & $-$0.142 \\
\hline
\end{tabular}
\caption{\label{tabH2L6}
Single number signatures of $H_{2}, L = 6$  for parameter values corresponding to the plots.}
\end{table}
%%%%%%%%%%%%%%%%%%%%%%%%%%%%%%%%%%%%%%%%%%%%%%%%%%%%%
\begin{table}[!tbp]
%\begin{ruledtabular}
\begin{tabular}{ccc}
\hline
 $\gamma$ & $\langle r \rangle$ & $-\langle \cos \theta \rangle$ \\
\hline
0.01 & 0.559  & 0.219  \\
% 0.5 & 0.695818 &  0.002376 \\
% 1.2 & 0.707291 & 0.056992 \\
2.1& 0.714&  0.062  \\
\hline
 $\lambda$ & $\langle r \rangle$ & $-\langle \cos \theta \rangle$ \\
\hline
0.01 & 0.675  & 0.006 \\
% 0.5 & 0.704122  & 0.065318 \\
1.2 & 0.715 & 0.080 \\
% 2.1 & 0.695234 & $-$0.006187 \\
\hline
 $\lambda_{1}$ & $\langle r \rangle$ & $-\langle \cos \theta \rangle$ \\
 \hline
 0.001 & 0.659 & 0.019 \\
0.5 & 0.704 & 0.057 \\
% 1.2 & 0.692817 & 0.000458 \\
% 4 & 0.686929 & $-$0.038638\\
\hline
\end{tabular}
\caption{\label{tabH2L8}
Single number signatures of the $H_{2}$, $L = 8$, for parameter values corresponding to the plots.}
\end{table}
%%%%%%%%%%%%%%%%%%%%%%%%%%%%%%%%%%%%%%%%%%%%%%%%%%%%%
\subsection{Plots for $H_{3}$}

\begin{figure*}[!tbp]
\centering
\includegraphics[width=15cm]{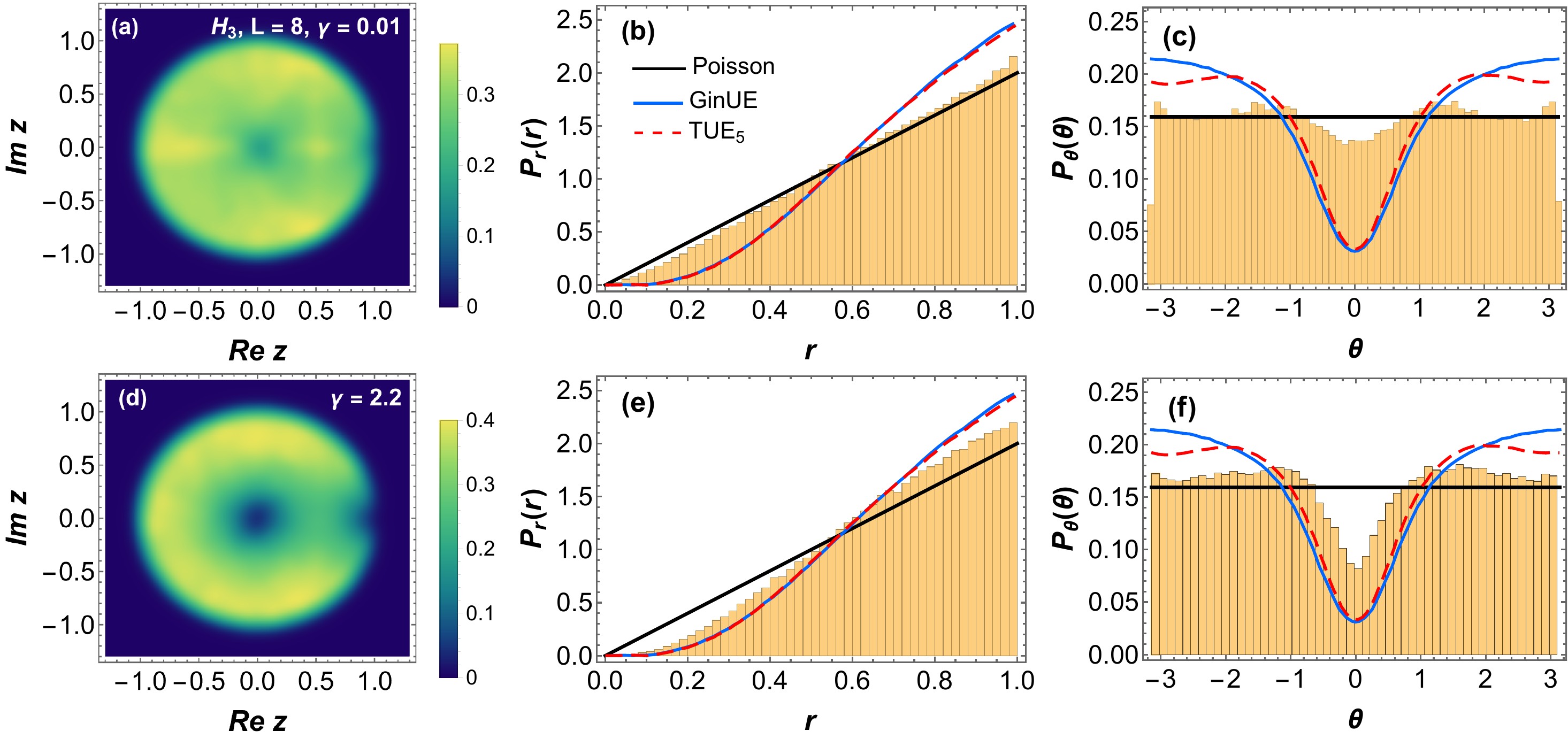}
\caption{Same plots for Fig.~\ref{H2-L6-gamvar} but for $L = 8$ case of the $H_{3}$ hamiltonian. $\gamma$ = 0.01 in (a)-(c) and 2.2 in (d)-(f). Close-to-GinUE behaviour is observed for $\gamma = 2.2$, also evident from the marginal distributions which are close to TUE$_{5}$ results plotted with the red dashed line. For $\gamma = 0.01$, results distinct from either Poisson or GinUE-like statistics is observed from a bow-arrow like structure embedded in the unit circle for the density of CSR. The corresponding radial and angular show noticeable deviation from the uniform distribution of the Poisson statistics.}
\label{H3-L8-gamm}
\end{figure*}

\begin{figure*}[!tbp]
\centering
\includegraphics[width=15cm]{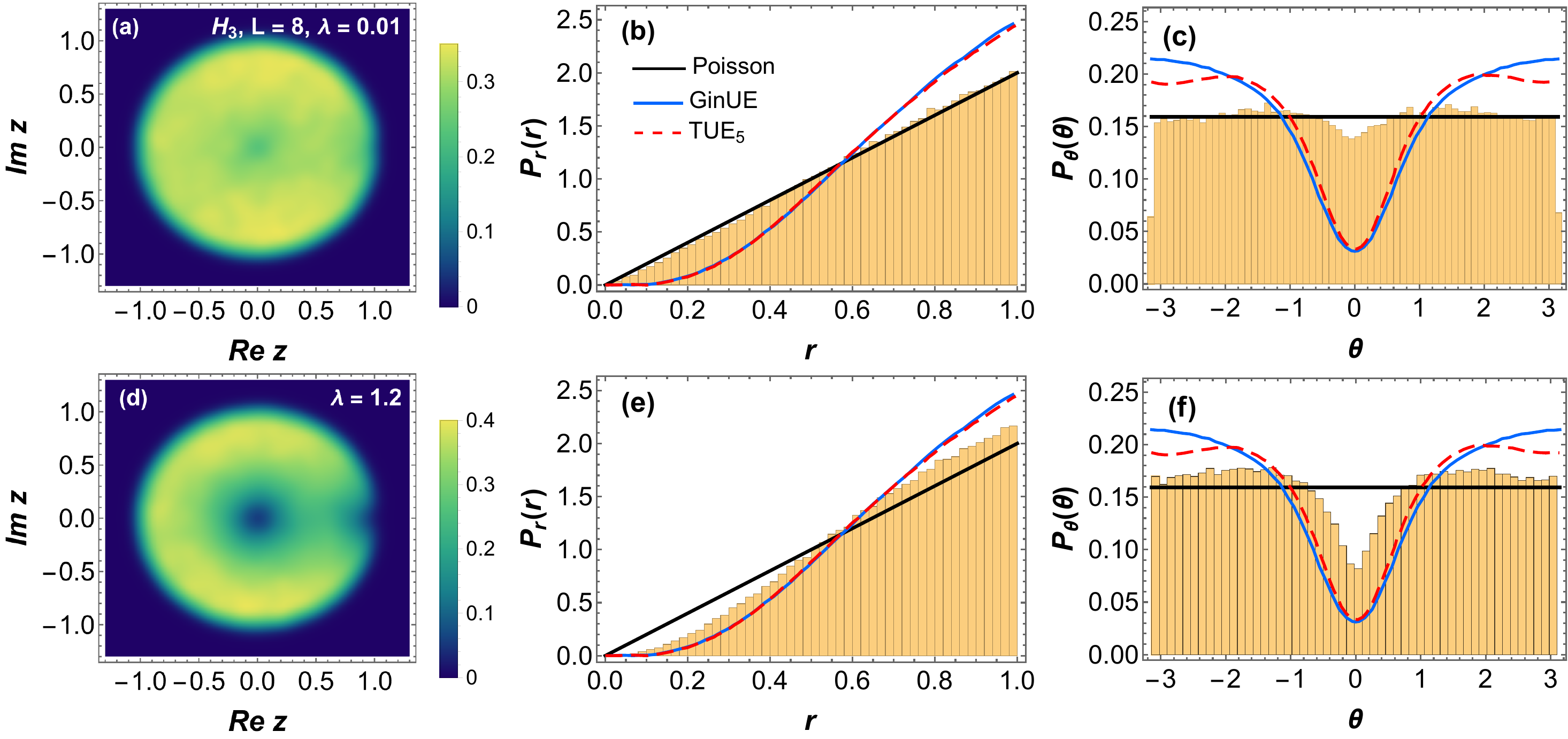}
\caption{In this plot $\lambda$ is varied for $L = 8$ case of $H_{3}$. Poisson-like behaviour is observed for $\lambda = 0.01$ with some deviations in the case of the angular distribution while $\mathrm{TUE}_{5}$-like behaviour is observed for $\lambda = 1.2$, also evident from the angular marginal distributions but in this case the radial distribution varies slightly from the analytical results.}
\label{H3-L8-lamb}
\end{figure*}

\begin{figure*}[!tbp]
\centering
\includegraphics[width=15cm]{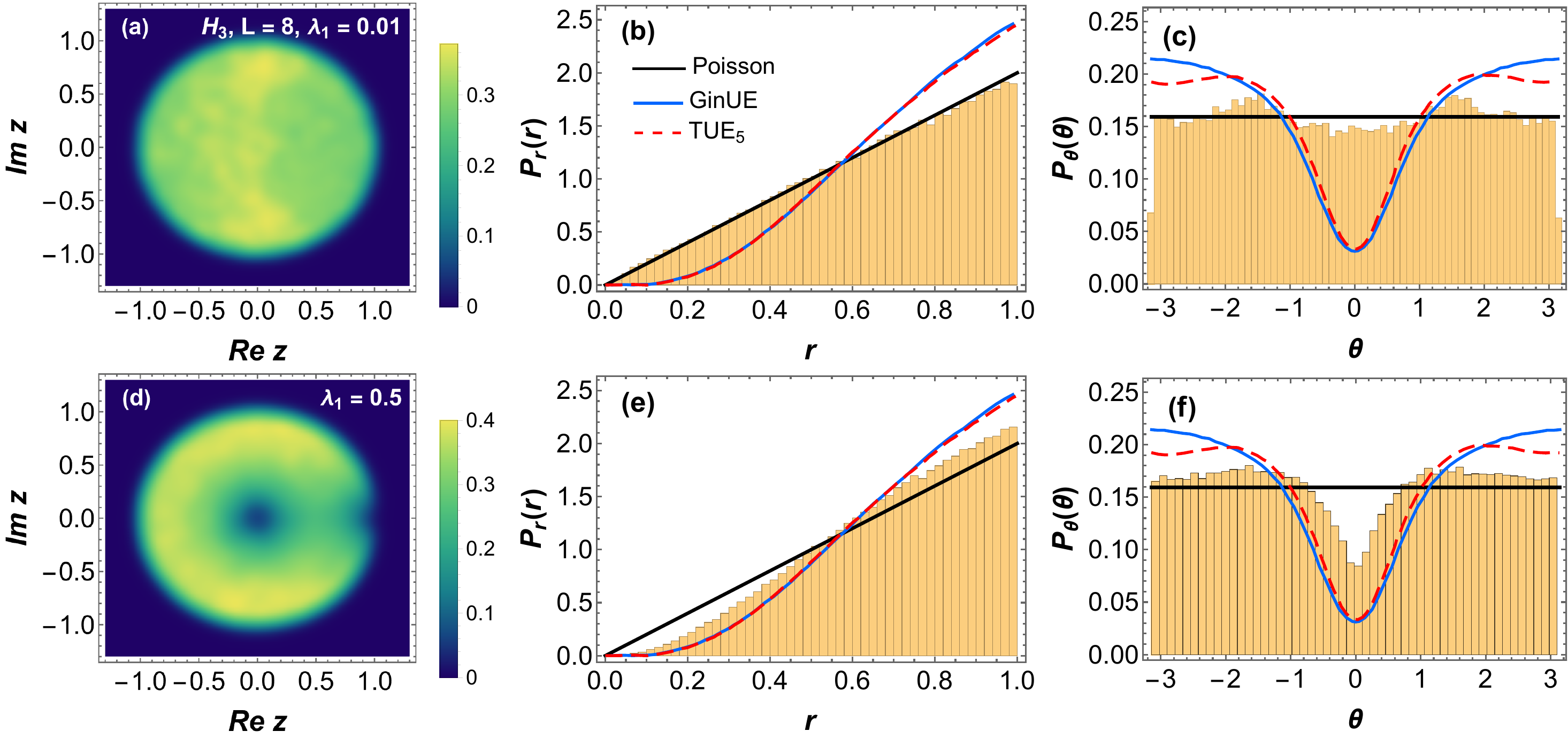}
\caption{In this case $\lambda_{1}$ is varied for $L = 8$ case of $H_{3}$. For $\lambda_{1} = 0.01$, Poisson-like statistics is observed with some deviations near the origin for the angular marginal distribution while the radial distribution matches the analytical results very well. For $\lambda_{1} = 0.5$ in (d)-(f), subtle expressions of $\mathrm{TUE}_{5}$ statistics is noted with some deviations for the radial distribution.}
\label{H3-L8-lamb1} 
\end{figure*}

For $H_{3}$,  we inspect the spectral properties for a chain length of $L = 8$ in Figs.~\ref{H3-L8-gamm}, \ref{H3-L8-lamb} and \ref{H3-L8-lamb1}.  In Fig.~\ref{H3-L8-gamm}, for $\gamma=0.01$, the distribution of CSR is quite distinct compared to the other spin chains or the matrix models. Due to the imaginary transverse field, not only do we have a significant number of complex eigenvalues, but the presence of complex-conjugate pairs gives rise to the bow-arrow like structure embedded in the unit circle of the density of CSR plot. We also see increased brightness along the real line suggesting the presence of real eigenvalues of the type $\pm x$. For the corresponding radial distribution we see deviation from Poisson-like statistics for $r$ values close to 1. The angular distribution too shows distinct peaks at the origin and at $\theta$ values of $\pm \pi$, thus deviating from the expected uniform distribution of the Poisson. The spectral statistics of $H_{3}$ when $\gamma$ is small is neither close to MM1 or MM2. However prominent GinUE-like features appear for $\gamma = 2.2$. Also we note that non-integrable behaviour persists for larger values of $\gamma$, compared to $H_{2}$. As expected on increasing the strength of the parameters, the results deviate significantly from RMT or Poissonian statistics. On the other hand for the variation of $\lambda$ and $\lambda_{1}$, the numerical results are captured very well by those from MM2, interpolating between 2D-Poisson and GinUE. This becomes evident by comparing the plots in Figs.~\ref{PC-GinUE-MM2-N256} or \ref{PC-GinUE-MM2-N2000} with \ref{H3-L8-lamb} and \ref{H3-L8-lamb1}. The exceptional cases which arise for the $\gamma$ variation in Fig.~\ref{H3-L8-gamm} has already been discussed above. The single number signatures for $H_3$ are given in Table~\ref{tabH3L8} and for certain values of the system parameters the results are quite close to GinUE (cf. $\langle r \rangle \sim$ 0.711 for $\gamma = 2.2$, 0.707 for $\lambda = 1.2$ and 0.705 for $\lambda_{1} = 0.5$). Negative values for $-\langle \cos \theta \rangle$ are also observed for $\lambda$ and $\lambda_{1}$ variation in this case.
%%%%%%%%%%%%%%%%%%%%%%%%%%%%%%%%%%%%%%%%%%%%%%%%%%%%%%%%%%%
%%%%%%%%%%%%%%%%%%%%%%%%%%%%%%%%%%%%%%%%%%%%%%%%%
%%%%%%%%%%%%%%%%%%%%%%%%%%%%%%%%%%%%%%%%%%%%%%%%%
%%%%%%%%%%%%%%%%%%%%%%%%%%%%%%%%%%%%%%%%%%%%%%%%%%%%%%
\begin{table}[!tbp]
%\begin{ruledtabular}
\begin{tabular}{ccc}
%\multicolumn{3}{|c|}{} \\
\hline
$\gamma$ & $\langle r \rangle$ & $-\langle \cos \theta \rangle$ \\
\hline
0.01 & 0.684 & 0.004   \\
% 0.1 & 0.688622 & 0.031650   \\
2.2 & 0.711 & 0.068 \\
% 4.2 & 0.703957 & 0.018581 \\
\hline
$\lambda$ & $\langle r \rangle$ & $-\langle \cos \theta \rangle$ \\
\hline
0.01& 0.675 & $-$0.007 \\
% 0.1 & 0.68667 & -0.024900\\
1.2 & 0.707 & $-$0.065 \\
% 3 & 0.680294 & 0.0773\\
\hline
$\lambda_{1}$ & $\langle r \rangle$ & $-\langle \cos \theta \rangle$ \\
\hline
0.01 & 0.662 & $-$0.015 \\
0.5 & 0.705 & $-$0.062  \\
% 2.1 & 0.688233 & -0.004395 \\
% 3.5 & 0.683496 &  0.039766\\
\hline
\end{tabular}
\caption{\label{tabH3L8}Single number signatures of the $H_{3}$, $L = 8$, corresponding to the plots shown previously.}
\end{table}
%%%%%%%%%%%%%%%%%%%%%%%%%%%%%%%%%%%%%%%%%%%%%%%%%%%%%%
\section{Summary}
\label{NH-sum}
In this paper we have studied the short range spectral fluctuation properties of three non-Hermitian spin chain models using complex spacing ratios, their marginal densities and the corresponding single number signatures. Although studies of integrability breaking and symmetry crossovers due to disorder, defects or random Zeeman fields are very common in Hermitian spin chain literature, they are more or less scarce for non-hermitian spin chains especially in the context of RMT. The key aspects considered in this paper can be summarised in the following three points.

Firstly, in these spin chains non-hermiticity has been rendered by the addition of complex coupling constants and imaginary random fields. Therefore they are not ``open" in the usual sense, since there is no direct interaction with the environment (which may be a larger spin system) such as in boundary driven spin chains. In fact it can be thought that the dissipative nature of these spin chains have been manifested through the imaginary system parameters.

Secondly, the symmetry properties of these hamiltonians, the breaking of which makes them undergo symmetry class transitions, makes them compelling case studies in the milieu of RMT.  For example, the anisotropic non-hermitian XY model with transverse field, i.e. $H_{1}$ has been shown to be $\mathcal{RT}$ symmetric due to its construction in Ref.~\cite{ZS2013} which plays the same role as $\mathcal{PT}$-symmetry in other pseudo-hermitian spin chain models. We have modified $H_{1}$ by the addition of an extra random longitudinal field in the case of $H_{2}$ and for $H_{3}$, the $z$-field has been made imaginary keeping the perturbative longitudinal $x$-field intact. These modifications lead to the breaking of their $\mathcal{RT}$ invariance which is turn facilitates integrability breaking and a transition from Poisson to GinUE-resembling statistics of RMT is observed on fine tuning of system parameters. The single number signatures are also quite close to GinUE ($\sim 0.74$) in several cases discussed previously. It is expected that these signatures of non-integrability and overlap with RMT improve further with increasing system dimension. One of the constraints in examining larger system dimensions in cases like this is the ever-increasing Hilbert space for full space diagonalization which requires greater computational resources and time. Despite this, a more thorough exploration is one of our future ventures.

Thirdly, we also provide simulation results for phenomenological matrix models of 1D and 2D Poisson to GinUE crossovers which in most scenarios approximate the spectral transitions shown by these spin chains rather well. These kind of interpolating models are already well known for Poisson-GOE and Poisson-semi-Poisson distributions. The first matrix model capturing 1D-Poisson to GinUE crossover is particularly interesting in the context of real to complex eigenvalue-kind of transitions which have been discussed in the context of many-body localization transition in non-Hermitian systems. This transition has previously been used as a diagnostic tool to understand many-body localization in hard-core boson models~\cite{HKU2019}. It will be of interest to see whether this same diagnostic can be used to understand ergodicity and many-body localization in our spin chain models. Lastly an intriguing but challenging problem would be to deduce analytical results for CSR in crossover matrix models of the kind considered in this paper.

\acknowledgements
AS acknowledges the research fellowship from DST-INSPIRE (IF170612), Govt of India and expresses gratitude to Prof. Manas Kulkarni with whom she had fruitful discussions on the subject while participating in the program - Statistical Physics of Complex Systems at ICTS, Bangalore (code: ICTS/SPCS2022/12).
SK acknowledges the support provided by SERB, DST, Govt. of India (Grant No. CRG/2022/001751).
%%%%%%%%%%%%%%%%%%%%%%%%%%%%%%%%%%%%%%%%%%%%%%%%%%%%%%%%%%%%%%%

\end{document}